\documentclass[12pt]{amsart}
\usepackage{amsmath}
\usepackage{color}
\usepackage{amsfonts}
\usepackage{amssymb}
\usepackage{epsfig}
\usepackage{times}
\usepackage{braket}
\usepackage{shadow}
\usepackage{fancybox}
\usepackage{colordvi}
\usepackage{graphics}
\usepackage{epsf,graphics,mathrsfs,yfonts,eufrak,simplewick}
\usepackage{psfrag}
 \usepackage[enableskew]{youngtab}       

\newcommand{\be}{\begin{eqnarray*}}
\newcommand{\en}{\end{eqnarray*}}
\newcommand{\ba}{\begin{array}}
\newcommand{\ea}{\end{array}}
\newcommand{\bea}{\begin{eqnarray}}
\newcommand{\ena}{\end{eqnarray}}
\def\bel{\begin{eqnarray}}
\def\enl{\end{eqnarray}}
\def\({\left(}
\def\){\right)}
\def\[{\left[}
\def\]{\right]}
\newcommand{\half}{\textstyle{\frac 1 2 }}

\newcommand{\nn}{\nonumber}
\newcommand{\slt}{\mathfrak{sl}_2}
\newcommand{\slth}{\widehat{\mathfrak{sl}}_2}

\newcommand{\ad}{{\rm ad}}

\newcommand{\Tr}{{\rm Tr}}

\newcommand{\la}{\lambda}

\newcommand{\al}{\alpha}

\newcommand{\s}{\sigma}
\newcommand{\z}{\zeta}

\begin{document}

\begin{title}[Exact density matrix for quantum group invariant sector of XXZ model]{
Exact density matrix for quantum group invariant sector of XXZ model}

\end{title}

\author{  F.~Smirnov${}^1$}\footnote{Membre du CNRS}

\address{
 Sorbonne Universit\'e, UPMC Univ Paris 06\\ CNRS, UMR 7589, LPTHE\\F-75005, Paris, France}\email{smirnov@lpthe.jussieu.fr}

\dedicatory{Dedicated to Nikolai Reshetikhin on the upcoming occasion of 
his  sixties birthday}

\begin{abstract}
Using the fermionic basis we obtain the expectation values of all 
$U_q(\slt)$-invariant 
local operators
on 8 sites for the anisotropic six-vertex model on a cylinder with generic Matsubara 
data. 
In the case when the $U_q(\slt)$ symmetry is not broken this computation
is equivalent to finding the entire density matrix up to 8 sites.
As application, we compute the entanglement entropy without and with 
temperature, and compare the results with CFT predictions. 
 \end{abstract}

\maketitle

\section{Introduction}

This paper is dedicated to my longtime friend Nikolai Reshetikhin, and its main idea has much
in common with our join paper  \cite{RS}.
In this paper it was shown that the restriction of the degrees of freedom
for the scattering states of the sine-Gordon model with rational coupling constant is
a non-violent procedure: if the quantum group invariant operators are considered
the contributions form the states which do not satisfy the RSOS restriction vanish 
in the correlation functions. In the present paper we apply the same logic to 
rather different situation. To be precise we are talking here about two different quantum groups which 
can be combined into the modular double \cite{LF}.

Consider the XXZ spin chain in critical regime:
\begin{eqnarray}
\mathcal{H}=\textstyle{\frac{1}{2}}\sum\limits_{k=-\infty}^{\infty}
\left( 
\sigma_{k}^1\sigma_{k+1}^1+
\sigma_{k}^2\sigma_{k+1}^2+
\Delta\sigma_{k}^3\sigma_{k+1}^3
\right), \quad \Delta =\half(q+q^{-1})\,.
\label{Ham}
\end{eqnarray}
We often use the parameter $\nu$:
$$q=e^{\pi i\nu}\,.$$
It is well-known that the XXZ model is closely related to the quantum affine group $U_q(\slth)$. 
We consider one of its finite-dimensional subgroups $U_q(\slt)$.
For finite temperature consider  the modified partition function an correlation functions
$$Z=\Tr\(e^{-\frac{H}{T}} q^{-2S}\)\,\quad \langle O\rangle_T=\frac 1 Z \Tr\(e^{-\frac{H}{T}} q^{-2S}O\)\,,$$
where $S$ is the total spin: $S=\frac1 2 \sum\sigma^3_j$. The insertion of $q^{-2S}$ is important:
it corresponds to the ``quantum group invariant" trace
(see \cite{alg} for relevant discussion). In some sense we have a generalisation of
the Witten index used in the SUSY models. 

In fact we can consider more general case of generalised Gibbs distribution, in other words
we use rather arbitrary Matsubara data
(six-vertex model on a cylinder). Really crucial
property for our construction is that the Matsubara maximal vector is quantum group invariant
(we call this unbroken quantum group symmetry).
We explain this in Section \ref{unbroken} . 
The similarity with \cite {RS} is in the fact that for $q^r=1$ the states of the lattice model which
do not satisfy the RSOS restriction do not contribute. 

If the quantum group symmetry is unbroken, only the quantum group invariant operators 
possess non-vanishing expectation values, this is similar to the usual Lie group symmetry. 
We consider a finite interval $[1,n]$ of the lattice and introduce the density matrix $\rho$. Then for any local
operator $O$ localised on this interval we have
$$\langle O\rangle=\Tr\(q^{-\sum_{j=1}^n\sigma^3_j  }\ \rho \ O  \)\,.$$
Again, we use the invariant trace.

For 
$$\nu=1-\frac{r}{s}$$
the scaling theory of the lattice model obtained in this way must coincide  with the minimal
model $M_{r,s}$. In particular, $\nu=1/s$ correspond to unitary models, it can be shown that for them
all the contributions to the modified partition function are positive. Otherwise we have non-unitary models.

In the present paper we compute the density matrices up to $8$ sites. 
In order to check the agreement with the scaling limit we compute the Von Neumann entropy. For the non-unitary case
it has strange features: it may be negative getting more negative with temperature. This is not
very surprising in this case. Nevertheless we always find very good agreement with the CFT predictions \cite{cft,korepin,cardy}.

Na\"ive idea is that for rational $\nu$ the restricted case of the six-vertex model on a cylinder
coincides completely
with the RSOS case.
This, however, is not quite simple as we explain in Section \ref{SOS} . Presumably, for that reason our results differ from 
those of papers \cite{RSOS1,RSOS2}: they agree with the CFT formulae for usual, not effective, central charge and
depend continuously on $\nu$.

Our procedure is the same as in \cite{ms}: we compute the expectation values 
of invariant operators 
for arbitrary
Matsubara data and arbitrary twist $q^{\kappa S}$. However,  unbroken quantum
group symmetry is possible  only for $\kappa=-1$. For broken quantum group symmetry 
we don not obtain complete density matrix.
Considering all the operators, not only the invariant ones, is possible. We do not do it for two
reasons: first, it is technically more complicated, second, we find it more interesting to 
have in the scaling limit the central charge $c=1-6\nu^2/(1-\nu)$ then $c=1$ independently
of the coupling constant.

The paper is organised as follows. In Section 2 we briefly recall some definitions concerning
the expectation value on a cylinder. Section 3 contains useful for us information from the
theory of quantum groups. Section 4  gives an account of our computational procedure.
In Section 5 we give general explanations regarding the unbroken quantum group symmetry.
In Sections 6,7 we expose our numerical results and comparison to the CFT
for the cases of zero and non-zero temperature respectively.

\section{Matsubara expectation values}\label{general}

Consider the quantum affine group $U_q(\slth)$ and its universal $R$-matrix $\mathcal{R}$.
We use usual Cartan generators. 
Central charge  equals to $0$, so $h_1=-h_0$, denote $H=h_1$.
The affine quantum group allows evaluation representations $\pi ^{2s}_\z$ with $s$ being a spin, and $\z$ an evaluation
parameter. Fix two integers $n,L$ and define two representations:
$$\pi _\mathbf{S}=(\pi ^{1}_1)^{\otimes n},\quad \pi_\mathbf{M}=\pi_{\tau_0}^{2s_0}\otimes\cdots\otimes
\pi_{\tau_{L-1}}^{2s_{L-1}}\,.$$
with spins $s_j$ and inhomogeneities $\tau_j$.
Later we shall use
$$t_j=\tau_j^2\,.$$
The index $\mathbf{S}$ stands for ``space" and $\mathbf{M}$ stands for Matsubara. 

We have the image of the universal $R$-matrix:
$$T_{\mathbf{S},\mathbf{M}}=(\pi _\mathbf{S}\otimes \pi_\mathbf{M})(\mathcal{R})\,.$$
Consider further the commuting family of Matsubara transfer-matrices
$$T_\mathbf{M}(\z,\kappa)=(\Tr\otimes I)\[(\pi^1_\z\otimes \pi_\mathbf{M})(\mathcal{R}(q^H\otimes I))\]\,,$$
and their eigenvector $|\Psi\rangle$.

The main object of our study is a linear functionals on operators
$O_\mathbf{S}$  acting on the representation space of $\pi_\mathbf{S}$:
\begin{align}
Z_\kappa\{O_\mathbf{S}\}=\frac{\langle\Psi| T_{\mathbf{S},\mathbf{M}}\ O_\mathbf{S}\ q^{2\kappa S_\mathbf{S}}|\Psi\rangle}
{\langle\Psi| T_{\mathbf{S},\mathbf{M}}\ q^{2\kappa S_\mathbf{S}}|\Psi\rangle}\,,\label{Z}
\end{align}
where $S_\mathbf{S}=1/2\sum_{j=1}^n\sigma^3_j$, in other words $2S_\mathbf{S}$ is the Cartan generator $H$ evaluated on $\pi_\mathbf{S}$.
If $O_\mathbf{S}$ is localised  on smaller interval that $[1,n]$ the space shrinks automatically. Clearly, $Z_\kappa$ can be
considered as such automatic reduction for the case $\mathbf{S}=(-\infty,\infty)$ if $|\Psi\rangle$ is the eigenvector with
maximal in absolute value eigenvalue of $T_\mathbf{M}(1)$. This explains importance of $Z_\kappa$ for physical
applications. 
The main tool of computation is the fermionic basis.

\section{Quantum group invariant operators}\label{QG}

\subsection{Generalities} Let us start with some simple facts concerning the quantum groups. 
We use Drinfeld's notations \cite{drinfeld}. Take a  basis  $e_a$, $e_0=1$. The comultiplication is 
a homomorphism
$$\Delta(e_a)=\mu_{a}^{bc}e_b\otimes e_c\,.$$
We have the $R$-matrix
\begin{align}\mathcal{R}\mu_{a}^{bc}e_b\otimes e_c=\mu_{a}^{bc}e_c\otimes e_b\mathcal{R}\,,\label{Rmat}\end{align}
and the antipode
$$\mu_{a}^{bc}s(e_b)e_c=\mu_{a}^{bc}e_bs(e_c)=\mu_{a}^{bc}s^{-1}(e_c)e_b=\mu_{a}^{bc}e_cs^{-1}(e_b)=\delta_{a,0}e_0\,,$$
which is an anti-automorphism. 
We define two adjoint actions
$$ad_{e_a}(x)=\mu_{a}^{bc}e_bxs(e_c)\,,\quad
ad^*_{e_a}(x)=\mu_{a}^{bc}s^{-1}(e_c)xe_b\,.$$
The first is an homomorphism, the second is an anti-homomorphism. 
In any quantum double $s^2$ is an internal homomorphism:
$$s^2(x)=YxY^{-1}\,,$$
for certain $Y$. 
The pairing
$$\langle x_1,x_2\rangle=\Tr(Yx_1x_2)\,,$$
is an invariant scalar product of operators
for $\Tr$ being any cyclic functional, for example usual trace over a finite-dimensional
representation if any. 
We have
$$\langle x_1,ad_{e_a}(x_2)\rangle=\langle ad^*_{e_a}(x_1),x_2\rangle\,.$$

Below we give some additional formulae which will be used only Section \label{unbroken}.
From now on we consider two representations, for economy of space we 
not distinguish between the generators of the quantum group and their representations, everything should be clear
from the context. 

Consider an invariant vector $v$. From \eqref{Rmat} one derives
\begin{align}
(I\otimes e_a)\mathcal{R}(I\otimes v)=\(ad^*_{e_a}\otimes I\)\mathcal{R}(I\otimes v)\,.
\label{ER}
\end{align}
Act by $O\otimes I$ 
on \eqref{ER} from the left, then take the invariant trace with respect to the first tensor
component obtaining
\begin{align}
&(I\otimes e_a)(\Tr\otimes I)\[\mathcal{R}(OY\otimes I)\](I\otimes v)=
(\Tr\otimes I)\[\(ad^*_{e_a}\otimes I\)(\mathcal{R})(OY\otimes I)\](I\otimes v)\label{adjoint}\\&=
(\Tr\otimes I)\[\mathcal{R}(ad_{e_a}(O)Y\otimes I)\](I\otimes v)\,.\nn\end{align}
In particular if $O$ is invariant
\begin{align}(I\otimes e_a)(\Tr\otimes I)\[\mathcal{R}(OY\otimes I)\](I\otimes v)=
\delta_{a,0}(\Tr\otimes I)\[\mathcal{R}(OY\otimes I)\](I\otimes v)\,.\label{inv}
\end{align}
So, acting on invariant vectors we obtain invariant ones. 

\subsection{Invariant operators}\label{IO}
The quantum group $U_q(\slth)$ contains two finite-dimensional subgroups isomorphic to
$U_q(\slt)$. Take one of them, for example the one created by $e_0,f_0,h_0$, which we denote respectively
by $E,F,-H$. We use $K=q^H$. 
The
defining relations of $U_q(\slt)$ are well-known, so, we give them without comments
just to fix the notations
$$
KE=q^2EK\,,\ \ KF=q^{-2}FK\,, \ \ [E,F]=\frac{K-K^{-1}}{q-q^{-1}}\,.
$$
The coproduct, counit and antipode are given by 
\begin{align}
&\Delta(E)=E\otimes 1+K\otimes E\,,
\quad \Delta(F)=F\otimes K^{-1}+1\otimes F\,,
\quad \Delta(K)=K\otimes K\,,\nn
\\
&\epsilon(E)=\epsilon(F)=0\,,\quad \epsilon(K)=1\,,\nn\\
&s(E)=-K^{-1}E\,,\quad s(F)=-FK\,,\quad s(K)=K^{-1}\,.\nn
\end{align}
Clearly,
$$Y=K^{-1}\,.$$

%%%%%%%%%%%%%%%%%
Consider an invariant under $U_q(\slt)$ operator $O_\mathbf{S}$. 
It is convenient to identify $O_\mathbf{S}$ with a vector $v$  in 
$V_1\otimes\cdots \otimes V_{2n}$, $V_k\simeq \mathbb{C}^2$
%$V_1\otimes\cdots \otimes V_n \otimes V_{\bar n}\otimes\cdots \otimes V_{\bar 1}$, $V_j\simeq \mathbb{C}^2$,  $V_{\bar j}\simeq \mathbb{C}^2$. 
The identification is 
\begin{align}
O_\mathbf{S}=\mathcal{I}(v)=v^{t_ {n+1},\cdots t_{2n}}c_{1}\cdots c_n\,, \label{ident}\end{align}
where $c_k$ acts from $V_{2n+1-k}$ to $V_k$ as
$$\quad c=\begin{pmatrix}0& 1 \\ -q^{-1}  &0\end{pmatrix}\,.
$$
For example for $n=1$ there is one invariant vector:
\begin{align}s_{1, 2}=-qe_{-1/2}\otimes e_{1/2}+e_{1/2}\otimes e_{-1/2}\,,\label{sing2}\end{align}
which under \eqref{ident} goes to the unit operator acting in $V_1$. 
Generally, we denote the element of the basis for an irreducible representation of spin $j$ by
$e_{-j},e_{-j+1},\cdots, e_j$.

We are going to give several definitions, for future convenience  they  are a bit more general than we need at this point. 
Bratteli diagram $J$ is a sequences $\{j_0,j_1,\cdots, j_k\}$ such that $j_{p+1}=j_p\pm1/2$, $j_p\ge 0$.
Lexicographically ordered totality of Bratteli diagrams (for fixed $k$) with $j=j_0$ and $j'=j_{k}$ will be denoted by
$B(k,j,j',\infty)$ , the meaning of the
latter argument will be clear later when we shall consider restricted case.
With every Bratteli diagram $J\in B(k,j,j',\infty)$ and $-j\le m\le j$, $-j'\le m'\le j'$ (integer of half-integer depending on $j$)
we associate a vector from $\(\mathbb{C}^2\)^{\otimes k}$:
$$
E_{k,J,m,m'}=\sum_{{\epsilon_1,\cdots \epsilon_{k}=\pm 1/2}\atop
m+\sum_{p=1}^k=m'}\ \ \prod_{p=1}^k
\begin{pmatrix}
j_{p-1}&1/2&j_{p}\\  m+\sum_{s=1}^{p-1}\epsilon_{s}&\epsilon_{p} &m+\sum_{s=1}^{p}\epsilon_s
\end{pmatrix}
e_{\epsilon_1}\otimes\cdots\otimes e_{\epsilon_{k}}\,.
$$
In order to avoid denominators which are dangerous for $q$ being a root of unity
we use non-normalised $3j$ symbols
\begin{align}
&\begin{pmatrix}
j&1/2&j+1/2\\m&-1/2 &m-1/2
\end{pmatrix}
=1\,,\quad \begin{pmatrix}
j&1/2&j+1/2\\m&1/2 &m+1/2
\end{pmatrix}
=\frac{q^{2(m+1)}-q^{-2j}}{q^2-1},\nn\\
&\begin{pmatrix}
j&1/2&j-1/2\\m&-1/2 &m-1/2
\end{pmatrix}
=1\,,\quad\begin{pmatrix}
j&1/2&j-1/2\\m&1/2 &m+1/2
\end{pmatrix}
=\frac{q^{2(m+1)}-q^{2(j+1)}}{q^2-1}\,.\nn
\end{align}
For $j=0$ this gives  an $U_q(\slt)$ decomposition of the tensor product of $k$ spaces $\mathbb{C}^2$:
\begin{align}
\(\mathbb{C}^2\)^{\otimes k}=\bigoplus\limits _{J\in B(k,0,j,\infty)}M_J\otimes V_j
\end{align}
 We have for the dimensions of multiplicity spaces
 $$\mathrm{card}(B(k,0,j,\infty))=\binom{k}{\frac{k-2j}2}-\binom{k}{\frac{k-2j}2-1}\,.$$

We shall use two bases of the space of invariant operators acting on  of $\(\mathbb{C}^2\)^{\otimes n}$.
The dimension of this space is the Catalan number $C_{n}$. 
%%%%%%%%%%%%%%%%%%%%%%%%%%%%%%%%%%%%%%%%%%%%%%%%%%%%
%We shall use two bases of invariant vectors in $V_1\otimes\cdots \otimes V_{2n}$
%(and consequently invariant operators acting in  $V_1\otimes\cdots \otimes V_{n}$ \eqref{ident}) both counted
%by Bratteli diagrams. Consider $k$ spaces $V_i\simeq  \mathbb{C}^2$. The multiplicities of the
%space of irreducible representation of spin $j$ are counted by sequences of $k$ half-integers
%$j_p$, such that $j_{p+1}=j_p\pm1/2$, $j_p\ge 0$, $j_1=1/2, j_k=j$. These are the Bratteli
%diagrams, we shall denote their ensemble by $B(k,j,\infty)$, the meaning of the
%latter argument will be clear later when we shall consider restricted case. We assume that
%$B(k,j,\infty)$ is lexicographically ordered.
%We have for the  cardinalities: 
%$$\mathrm{card}( B(k,j,\infty))=\binom{k}{\frac{k-2j}2}-\binom{k}{\frac{k-2j}2-1}\,.$$

1. To have an orthogonal basis we use the above construction with $3j$-symbols. 
The basis is given by 
$$O(J)=\mathcal{I}(E_{2n,J,0,0})\,.$$
This basis is   orthogonal, but not orthonormal
with respect to the scalar product
$$\langle O_1,O_2\rangle=\Tr\(q^{-\sum\sigma^3_j}O_1O_2\)\,.$$
The normalisation is
$$(O(J),O(J'))=\prod_{p=1}^{2n-1}\mathrm{dim}_q(j_p)\,,
$$
where the quantum dimension is
\begin{align}
\mathrm{dim}_q(j)=[2j+1]\,.\label{qdim}
\end{align}
Here and later
$$[k]=\frac{q^{k}-q^{-k}}{q-q^{-1}}\,.$$

2. 
For our computations it is convenient to use a simpler basis. 
Take $2n$ points on the real axis: $1,2,\cdots , 2n$ and connect them pairwise
by arcs in  the upper half plain requiring that the arcs do not intersect. 
For example,
\vskip .2cm
\centerline{\includegraphics[height=3cm]{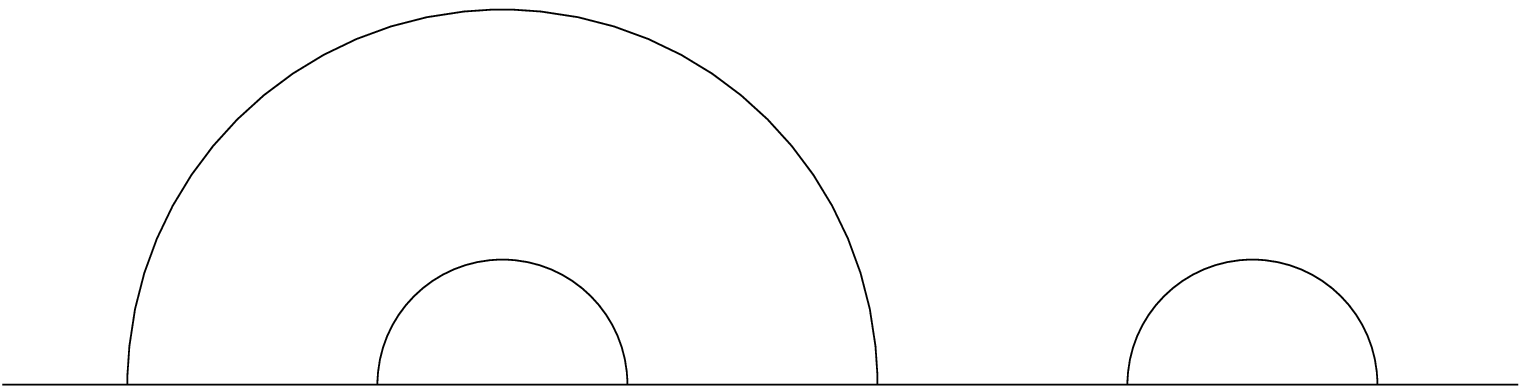}}
{\it fig.1} Simple construction of invariant vectors.
\vskip .2cm

With  any
such design associate the Bratelly diagram $J$  writing one half of the number of arcs
passing over every interval $[k,k+1]$. 
For instance, the {\it fig. 1} corresponds to $J=\{0,1/2,1,1/2,0,1/2,0\}$.
Denote by $i_1,\cdots, i_m$ the beginnings of the arcs, and
by $k_1,\cdots, k_n$ their ends. Then the vector associated with every design is
$$s_{i_1,k_1}s_{i_2,k_2}\cdots s_{i_n,k_n}\,,$$
with $s_{i,k}$ given by \eqref{sing2}, and
define
$$\widetilde{O}(J)=\mathcal{I}(s_{i_1,k_1}s_{i_2,k_2}\cdots s_{i_n,k_n})\,.$$
This is another basis.

Recall that $B(2n,0,0,\infty)$ is lexicographically ordered. We have two bases of invariant operators
$O(J)$ and $\widetilde{O}(J)$. It is easy to see that they are related by a triangular transformation
$$O(J)=U_{J,J'}\widetilde{O}(J')\,.$$
The  matrix $U$ is not hard to compute inductively.

Considering the interval $[1,n]$ we are not interested in the translationally
irreducible operators, i.e. the operators which are not  localised
on subintervals of smaller length. It is easy to figure out that the number
of translationally irreducible invariant operator is
$$ D_n=C_n-2C_{n-1}+C_{n-2}\,.$$
For the second basis eliminating the translationally reducible operators is
simple: it suffices to throw all the vectors containing $s_{1,2n}$ or $s_{n,n+1}$.

%%%%%%%%%%%%%%%%%%%%%%%%%%%%%%%%%%%%%%%%
\section{Procedure of computation}\label{compute}

Fermionic basis for the case of  $U_q(\slt)$-invariant  operators is parallel to
the $\slt$-invariant case for the XXX model which 
is explained in details in \cite{ms}. So, we shall be brief here. 
We have two sets of fermionic  operators $b_j,\ b^*_j$, $c_j,\ c_j^*$, ($j=1,2,3,\cdots$)
with canonical commutation relations,
and use notations  $b^*_{J}$, $c_{J}^*$ for products, $J$ being a strictly ordered
{multi-}index $\{j_1,\cdots,j_k\}$.
For two {multi-}indices of the same length
we write ${I}\preccurlyeq {J}$ if $i_p\le j_p$ for all $p$. We denote by {$|{I}|$} the sum of elements in ${I}$. 
Our fermionic  operators act on the space of local fields, role of vacuum is played by the unit
operator $\mathrm{I}$.
Consider the
space $\mathfrak{H}^{(n)}$ with the basis
\begin{align}b^*_{I^+}c_{I^-}^*\cdot \mathrm{I},\label{base}\end{align}
with $\#({I^+})=\#({{I^-}}){\le [n/2]}$, $\max({I^+}\cup {I^-})\le n$, 
${I^+}\preccurlyeq {I^-}$ (the difference with the XXX case is that we do not impose
$|{I^+}|+|{I^-}|\equiv 0\ (\mathrm{mod}\ 2)$, since there is no $C$-symmetry).
Define the operators
\begin{align}
&Q_m=\sum_{j=1}^{m-1}c_jb_{m-j}\,,\quad  2,3,\cdots\,;\qquad
M=\sum\limits_{i=1}^\infty c^*_ib_i\,.\nn
\end{align}
Introduce the space  $\widetilde{\mathfrak{H}}^{(n)}$ defined as above with the condition ${I^+}\preccurlyeq {I^-}$ lifted.
The operators $Q_m$ act from  $\mathfrak{H}^{(n)}$ to $\widetilde{\mathfrak{H}}^{(n)}$. The operator
$M$ acts from the space  $\widetilde{\mathfrak{H}}^{(n)}_2$ (space of charge $2$),  span by
the vectors \eqref{base} with $\#({I^+})+1=\#({{I^-}})-1{\le [n/2]}$, $\max({I^+}\cup {I^-})\le n$ , to
$\widetilde{\mathfrak{H}}^{(n)}$. We define the subspace 
${\mathfrak{V}}^{(n)}$ of ${\mathfrak{H}}^{(n)}$ 
by {
$$\mathfrak{V}^{(n)}=\{v\in\mathfrak{H}^{(n)}\mid Q_mv\in M\widetilde{\mathfrak{H}}^{(n)}_2
\ \text{for}\ \ m=n+1,n+2,\cdots\,\}.$$
}
It is easy to see that  $Q_m{\mathfrak{V}}^{(n)}=0$ for $m>2n-1$  , so the actual
number of requirements is finite.

Denoting basis of ${\mathfrak{V}}^{(n)}$ by $v_\al$ we have {$F=||F_{\al,\{{I^+},{I^-}\}}||$, the
first one} of several matrices used below:
$$v_\al=\sum\limits_{ {\#({I^+})=\#({{I^-}})}\atop{ \max({I^+}\cup {I^-})\le n,\ {I^+}\preccurlyeq {I^-}}}F_{\al,\{{I^+},{I^-}\}}\ b^*_{I^+}c_{I^-}^*\cdot \mathrm{I}\,.$$

Our goal is to find an analogue of OPE:
\begin{align}
\widehat{O}(J)=\sum\limits_{\al}c_{J,\al}(n)v_\al\,,\label{main}
\end{align}
where $\widetilde{O}(J)$ is an $U_q(\slt)$ invariant operator constructed via the Bratteli diagram $J$
and the second of the bases above.

As in the previous paper \cite{ms} we fix the OPE coefficients considering
finite Matsubara chains. For every Matsubara data we have an equation
\begin{align}
\langle \widehat{O}(J) \rangle=\sum\limits_{\al}c_{J,\al}(n)\langle v_\al\rangle \,.\label{eqs}
\end{align}

To compute the right hand side we use
$$\langle b^*_{I^+}c_{I^-}^*\cdot \mathrm{I}\rangle =\omega_{I^+,I^-}\,,$$
here and later
$$\omega_{I^+,I^-}=\det \bigl(\omega _{i^+_p,i^-_q}\bigr)_{p,q=1,\cdots, |I^+|}\,,$$
with $\omega_{i,j}$ are the Taylor series coefficients 
$$\omega(x,y)=\sum\limits _{i,j=1}^\infty\omega_{i,j}(x-1)^{i-1}(y-1)^{j-1}\,,$$
of  a function $\omega(x,y)$ is defined below. 
Here and later the latin letters are are squares of  the evaluation parameters of $U_q(\slth)$ representations. 

%%%%%%%%%%%%%%%%%%%%%%%%%%%%%%%%%%%%%%%%
With every eigenvector of the Matsubara transfer-matrix we associate an eigenvalue of the $Q$-operator
$$Q(x)=\prod\limits_{j=1}^m(1-x/b_j)\,,$$
The information about the Matsubara spin chain encoded is two functions
$$a(x)=x^L+\sum\limits_{j=1}^La_jx^{L-j}\,,,\quad d(x)=x^L+\sum\limits_{j=1}^Ld_jx^{L-j}\,.$$

We have the Bethe equations
$$q^{-m}a(b_j)Q(b_j q^2)+q^{m}d(b_j)Q(b_j q^{-2})=0,\quad j=1,\cdots m\,.$$
In principle we could introduce a twist  multiplying $a(x)$ and $d(x)$
by  $q^{\kappa}$ and  $q^{-\kappa}$ respectively, but 
our goal will be to obtain equations for the OPE coefficients, and practice shows that
twist does not produce independent ones.

Our main trick is to take for the input data
$$\{b_1,\cdots,b_m,a_{m+1},\cdots ,a_L,d_1,\cdots ,d_L
\}\,.$$
Then for the unknowns $a_1,\cdots ,a _m$ we have linear equations.

Introduce the measure
$$dm(x)=\frac{1}{1+\mathfrak{a}(x)}\frac{dx}{x},\quad \mathfrak{a}(x)=q^{-2m}\frac{Q(b_j q^2)}{Q(b_j q^{-2})}
\,,$$
the auxiliary function
$$\psi(x)=\frac{x+1}{2(x-1)}\,,$$
and kernels
$$K(x)=\psi(x q^{2})-\psi(x q^{-2})\,,\quad f_R(x)=\psi(x q^{-2})-\psi(x )
\,,\quad f_L(x)=\psi(x q^{2})-\psi(x )\,.$$
We have ``integral" equation
\begin{align}
G(x,z)=f_R(x,z)-\frac 1 {2\pi i}\oint_\Gamma K(x/y)G(y,z)dm(y)\,,
\end{align}
with $\gamma$ going around $y=b_1,\cdots , b_m,z$. For finite $m$ this is equivalent
to a system of linear equations for $m$ functions $G(b_j,z)$.
The function $\omega(x,y)$ \cite{bgks} is
\begin{align}
\omega(x,y)=-\frac 1 {2\pi i}\oint_{\Gamma '} f_L(x/y)G(y,z)dm(y)-\frac1 4 K(x/z)\,,\nn
\end{align}
where $\Gamma'$ encircles, in addition to $\Gamma$, the point $y=x$.

The expectation values of invariant operators  is computed 
exactly as in \cite{ms}, basically we rewrite in the basis of Young diagrams formulae of
\cite{FST,BIK,Slavnov}.

We repeat some definitions. 
Consider Young diagrams
{$Y_\la$ where $\la=(\la_1,\cdots,\la_n)$, $\la_i\ge\la_{i+1}>0$
is a partition. We set $\#(\la)=n$.}
{It is called the length of $Y_\la$.}
We work in the space $H_q$  whose elements are
$${Y=\sum_{\#(\la)\le q}c_\la Y_\la\,.}$$
{In the below we will identify $Y_\la$ with $\la$. The symbol $\emptyset$ denotes
the empty diagram.}
Define the operation {$\mathrm{cut}_q$ which acts} from $H_{q'}$ with $q'>q$ to $H_{q}$ erasing
all the terms with $\#(\la)>q$. 
Consider the Grassmann space $F_q$ with the 
basis $\psi^*_{k_1}\cdots \psi^*_{k_q}$ ($k_1>\cdots>k_q\ge 0$) .
{We have the usual isomorphism between the spaces $H_q$ and $F_q$.}
\begin{align}
&{\psi^*_{k_1}\psi^*_{k_2}\cdots\psi^*_{k_q}\ \mapsto\ (k_1-(q-1),k_2-(q-2),\cdots ,k_q)_0\,,}\label{maps}\\
&{(\la_1,\cdots,\la_{{n}})\ \mapsto\ \psi^*_{\la_1+q-1}\cdots\psi^*_{\la_{{n}}+q-{{n}}}\psi^*_{q-{{n}}-1}\cdots\psi^*_0}\,{,\ \text{where} \ n\leq q.}\nn
\end{align}
{In the above, $()_0$ }means removing all entries equal to $0$. 
Schur polynomial $s_\la(x_1,\cdots,x_q)$  is the symmetric polynomial
{
$$s_\la(x_1,\cdots,x_q)=\frac{\det ||x_j^{{\la_i+q-i}}||}{\det ||x_j^{q-i}||}\,.$$
The above formula gives an isomorphism between $H_q$ and $P_q$.}

For a given polynomial of one variable $P(x)=\sum_{j=0}^d p_j x^j$ we define
the operator $P\wedge \ F_{q-1}\subset F_q$ multiplying by $\sum_{j=0}^d p_j \psi^*_j$, this operator is
defined {as $P\wedge H_{q-1}\subset H_q$} by the isomorphism \eqref{maps}. 
This definition can be generalised to polynomials of several variables in obvious way. Certainly the polynomials anti-symmetrise themselves automatically. 

We shall also need the simplest Littlewood-Richardson formula for multiplication of a Schur polynomial by
elementary symmetric function {$\s_j$,} which translates as action on $H_q$
\begin{align*}
\sigma_j\circ(\la_1,\cdots, \la _{{n}})
=\sum\limits _{I^-}^{\left({{n}}+
\min(j,q-{{n}})\atop j\right)}
\Bigl((\la_1,\cdots, \la _{{n}},\underbrace{0,\cdots,0}_{\min(j,q-{{n}})})+e_{I^-}\Bigr)_{\mathrm{order}}\,,
\end{align*}
where $e_{I^-}$ are all vectors of dimension ${{n}}+\min(j,q-{{n}})$
with $j$ elements equal to $1$ other elements being $0$, ``order" means that 
we have to drop all the tables in which elements happen to be not ordered, {and}
we also drop all zeros in the final table.

In what follows we shall also need the operation $\mathrm{cut}_m({Y})$
which erases all the Young diagrams in ${Y}$ with lengths greater than $m$.

For a  partition $\la$ define the coefficients $e_{\la,\la'}$ via
$$s_\la(x_1,\cdots, x_k,1)=\sum\limits _{\la'}e_{\la,\la'}\ s_{\la'}(x_1,\cdots, x_k)\,.$$
This gives rise to an operator
$$EY_\la=\sum\limits _{\la'}e_{\la,\la'}Y_{\la'}\,.$$
The easiest  algorithm for finding  $e_{\la,\la'}$ consists in the following: 
express $s_\la(x_1,\cdots, x_k,1)$ through Jacobi-Trudi formula using 
the transposed $\la$. This formula is given in terms of the elementary symmetric
functions $\sigma_k(x_1,\cdots, x_k,1)$. Using
$$\sigma_k(x_1,\cdots, x_k,1)=\sigma_k(x_1,\cdots, x_k)+\sigma_{k-1}(x_1,\cdots, x_k)\,,$$
expand the Jacobi-Trudi  determinant getting a sum of Schur polynomials. 

Slavnov formula for the scalar product of on-shell and off-shell Bethe vectors 
$\langle b_1,\cdots, b_m|x_1,\cdots x_m\rangle$ is a symmetric
polynomial of the off-shell $x_1,\cdots x_m$ with the following representation in terms of the
Young diagrams:
\begin{align}\mathcal{N}(b_1,\cdots,b_m)=
\frac{\prod \bigl(b_j^{m-2}d(b_j)\bigr)q^{m(m-2)}(-1)^{\frac 1 2 m(m+1)}}{W(b_1,\cdots b_m)}
\Bigl(P_1\wedge\cdots\wedge P_m\wedge\emptyset\Bigr)\,,\nn
\end{align}
where
\begin{align}
P_j(x)=\frac {xb_j^2(q^2-1)} {x-b_j}\Bigl(q^{2-2m}a(x)\frac{Q(xq^2)}{xq^2-b_j}-d(x)
\frac{Q(xq^{-2})}{xq^{-2}-b_j}\Bigr)\,.\nn
\end{align}

We shall need also the Gaudin formula for normalisation:
$$G(b_1,\cdots_,b_m)=(q-q^{-1})^m\prod_{j=1}^m
a(b_j)d(b_j)
\prod_{i\ne j}
\frac
{b_iq-b_jq^{-1}}
{b_i-b_j}
\det b_i\partial_{b_i}(\log\mathfrak{a}(b_j))\,.$$

As usual we consider the matrix elements $T_{i,j}$ of the Matsubara monodromy matrix. Their
action in our framework translates into the the action of the operators $\mathcal{T}_{i,j}$
described below.

We begin with the operators $\mathcal{T}_{1,1}$, $\mathcal{T}_{2,2}$ which do not change the charge. 
\begin{align}
&\mathcal{T}_{1,1}Y=\sum_{k=0}^{l}\mathrm{cut}_{l}\Bigl(\sigma_k\circ E A_k\wedge Y
\Bigr)\,,\qquad
A_k(x)=(-1)^kq^{l-2k}x^{l-k}a(x)\,,\nn\\
&\mathcal{T}_{2,2}Y=\sum_{k=0}^{l}\mathrm{cut}_{l}\Bigl(\sigma_k\circ E D_k\wedge Y
\Bigr)\,,\qquad D_k(x)=(-1)^kq^{-l+2k}x^{l-k}d(x)\,.\nn
\end{align}
The most complicated operator is $\mathcal{T}_{1,2}$ which raises the number of variables:
\begin{align}
&\mathcal{T}_{1,2}Y=\sum_{j,k=0}^{l}\mathrm{cut}_{l}\Bigl(\sigma_j\circ\sigma_k\circ E B_{j,k}\wedge Y\Bigr)+\sum_{k=0}^{l}\mathrm{cut}_{l}\Bigl(\sigma_k\circ  B_k\wedge Y
\Bigr)\,,
\end{align}
where $B_{j,k}$, $B_{k}$ are  polynomials of respectively two and one variables:
\begin{align}
& B_{j,k}(x,y)=(-1)^{j+k}q d(y)a(x)(q^{-2}y)^{l-k}\frac {(xq^2)^{l-j}-y^{l-j}
 } {xq^2-y}\,,\nn\\
&B_{k}(x)=(-1)^{k+1}q^{-1}d(1)a(x)\frac {q^{-2(l-k)}-x^{l-k}} {xq^{-2}-x}
\,.\nn
\end{align}
Finally, the operator $\mathcal{T}_{2,1}$ which lowers the number of variables
is the simplest one:
$$\mathcal{T}_{2,1}Y=EY
\,.$$

In order to compute $\langle O\rangle$ we present $O$ as a sum of the 
operators
$E_{i_1,j_1}\cdots E_{i_n,j_n}$, and use
\begin{align}
\langle E_{i_1,j_1}\cdots E_{i_n,j_n}\rangle=\frac 1 {G(b_1,\cdots,b_m)}
\mathrm{Schur}_{b_1,\cdots, b_m}\Bigl(\mathcal{T}_{i_n,j_n}\cdots \mathcal{T}_{i_1,j_1}\mathcal{N}(b_1,\cdots,b_m)
\Bigr)\,,
\end{align}
where 
$\mathrm{Schur}_{b_1,\cdots, b_m}$ is a linear functional on the vector space $H_0$ 
which maps every Young diagram to corresponding Schur polynomial of arguments
$b_1,\cdots, b_m$. 
Now we can construct as many equations for the coefficients of OPE as we wish. 

The rest of the computations follow closely that of \cite{ms}. For example, for
$n=8$ we have 324 fermionic vectors $v_\al$, and construct 20 eqs. with $L=1,m=0$,
120 eqs. with $L=2,m=0$;
100 eqs. with $L=3,m=0$;
10 eqs. with $L=4,m=0$;
2 eqs. with $L=2,m=1$;
50 eqs. with $L=3,m=1$;
70 eqs. with $L=4,m=1$;
2 eqs. with $L=5,m=1$.
Then we proceed with Gauss triangularisation. This is easy to do for numeric value of $q$,
but rather impossible keeping $q$ as variable. That is why we would like
to proceed with interpolation. But the solutions for the coefficients
$c_{J,\al}(n)$ contain denominators which we have to fix, then interpolating for
the numerators is possible, but we also have to estimate the degree of them as functions of $q$. All these data can be guessed  considering several 
numerical examples.

We have for the lowest common denominators ($\mathrm{den}(n)$) and for the maximal exponent of the numerators  ($\mathrm{mn}(n)$):
\begin{align}
&\mathrm{den}(2)=4q,\quad\quad\quad\quad\quad\quad\quad\quad\quad\  \, \mathrm{mn}(2)=2\label{den}\\
&\mathrm{den}(3)=4q[2],\ \quad\quad\quad\quad\quad\quad\quad\ \ \ \ \,  \mathrm{mn}(3)=2\nn\\
&\mathrm{den}(4)=16q^5[2][3],\quad\quad\quad\quad\quad\quad\ \
  \mathrm{mn}(4)=10\nn\\
&\mathrm{den}(5)=16q^8[2][3][4],\quad\quad\quad\quad\quad\  \ \mathrm{mn}(5)=16\nn\\
&\mathrm{den}(6)=64q^{15}[2][3]^2[4][5],\quad\quad\quad\ \  \ \mathrm{mn}(6)=30\nn\\
&\mathrm{den}(7)=64q^{22}[3][4]^2[5][6],\quad\quad\quad\ \ \   \mathrm{mn}(7)=44\nn\\
&\mathrm{den}(8)=256q^{33}[3]^2[4]^2[5][6][7],\quad\ \ \ \, \mathrm{mn}(8)=74\nn
\end{align}

We proceed with interpolation obtaining finally matrices $c_{J,\al}(n)$.
Now for any Matsubara data  we have
\begin{align}
\langle O(J)\rangle=U_{J,J'}c_{J',\al}F_{\al,I^+,I^-}\omega_{I^+,I^-}\,.\label{main}
\end{align}
Here only $\omega_{I^+,I^-}$ depends on the Matsubara data.

Let us give an example. For $n=4$ the basis $v_\al$ consists of 9 elements
\begin{align}
&v_1=\mathrm{I}\,,\ v_2=b^*_1c^*_1\cdot\mathrm{I}\,,\ v_3=b^*_1c^*_2\cdot\mathrm{I}\,,\ v_4=b^*_1c^*_3\cdot\mathrm{I}\,,\ v_5=(b^*_1c^*_4-b^*_2c^*_3)\cdot\mathrm{I}\,,\ v_6=b^*_2c^*_2\cdot\mathrm{I}\,,\nn\\
&v_7=(b^*_2c^*_4-b^*_3c^*_3)\cdot\mathrm{I}\,,\ v_8=b^*_2b^*_1c^*_2c^*_1\cdot\mathrm{I}\,,\ v_9=(b^*_2b^*_1c^*_3c^*_2+b^*_2b^*_1c^*_4c^*_1-b^*_3b^*_1c^*_3c^*_1)\cdot\mathrm{I}\,.\nn
\end{align}
For $\widetilde{O}(J_1)$ with $J_1={\{1/2,0,1/2,0,1/2,0,1/2,0\}}$ which is constructed from
 $s_{1,2}s_{3,4}s_{5,6}s_{7,8}$ we have
\begin{align}
&\widetilde{O}(J_1)=\frac{(1 + q^2)^2}{(16 q^2)}v_1
+\frac{ (1 + q^4)(1 - q^2) (1 + 10 q^2 + q^4)}{
 4 q (1 + q^2) (1 - q ^6)}v_2
    +\frac{ (1 - q^2)^4 }{
  2 (1 + q^2) (1 - q ^6)}v_3\nn\\&+\frac{4 (1 - q^2)^3 q }{(1 + q^2) (1 - q ^6)}
 \bigl( -4v_4
    + 
    3v_6
    -v_7\bigr)
  +\frac{ 
  q^2(1 - q^2)}{(1 - q^6)}\bigl(2v_8-v_9\bigr)\,.\nn
\end{align}
By triangularity $O(J_1)=\widetilde{O}(J_1)$.
Certainly, for $n>4$ the formulae are getting more complicated, but their
structure is more inspiring that in XXX case.

\section{The case of unbroken quantum group symmetry}\label{unbroken}

Consider the functional $Z_\kappa$ for particular case $\kappa=-1$:
\begin{align}
Z_{-1}\{O_\mathbf{S}\}=\frac{\langle\Psi| T_{\mathbf{S},\mathbf{M}}\ O_\mathbf{S}\ q^{-2 S_\mathbf{S}}|\Psi\rangle}
{\langle\Psi| T_{\mathbf{S},\mathbf{M}}\ q^{-2 S_\mathbf{S}}|\Psi\rangle}\,.\label{Z}
\end{align}
From the identity \eqref{inv} one concludes that the transfer-matrix $T_\mathbf{M}(\z)=T_\mathbf{M}(\z,-1)$ preserves the
$U_q(\slt)$-invariant subspace of the Matsubara space. Hence there are invariant eigenvectors. Let $|\Psi\rangle$
by one of them:
$$\pi_\mathbf{M}(e_a)|\Psi\rangle=\delta_{a,0}|\Psi\rangle,\quad \langle\Psi|\pi_\mathbf{M}(e_a)=\delta_{a,0} \langle\Psi|\,.$$
With such choice we say that the quantum group symmetry is unbroken. Then we conclude from \eqref{adjoint} that
\begin{align}
&\delta_{a,0} \langle\Psi|\Tr_\mathbf{S}\[T_{\mathbf{S},\mathbf{M}}\ O_\mathbf{S}\ q^{-2 S_\mathbf{S}}\]|\Psi\rangle=\langle\Psi|\Tr_\mathbf{S}\[\pi_\mathbf{M}(e_a)T_{\mathbf{S},\mathbf{M}}\ O_\mathbf{S}\ q^{-2 S_\mathbf{S}}\]|\Psi\rangle
\nn\\&
=\langle\Psi|\Tr_\mathbf{S}\[T_{\mathbf{S},\mathbf{M}}\ \ad_{e_a}\(O_\mathbf{S}\)\ q^{-2S_\mathbf{S}}\]|\Psi\rangle\,.\nn
\end{align}
So, like in the case of invariance under usual Lie group in the unbroken case only invariant operators count.
Hence our computations give access to the entire density matrix matrix for the space interval $[1,n]$ with
$|\Psi\rangle$ describing the environment. Before going any further let us see what happens for $q$ being a root
of unity. To simplify notations let us assume that all Matsubara spins equal $1/2$. Certainly, in this case
$L$ must be even to have an invariant subspace. Construct the basis in the invariant subspace of the Matsubara
space using the second basis from the Section \ref{QG} (replacing $2n$ by $L$). The linear independence is
obvious for any $q\ne 0$. Construct the matrix of scalar products of the basis vectors and compute its
rank. If $q^r=1$ the rank is smaller than the maximal. This is well-known as well as the fact that
the rank equals cardinality of restricted Bratteli diagrams in which all the intermediate spins 
\begin{align}j_p\le\frac{r-2} 2\,.\label{restr}\end{align}
Since action of $\Tr_\mathbf{S}\[T_{\mathbf{S},\mathbf{M}}\ O_\mathbf{S}\ q^{-2 S_\mathbf{S}}\]$ for invariant $O_\mathbf{S}$
preserves the invariant subspace and at the end we take the scalar product with invariant vector $\langle\Psi|$
the states which do not satisfy \eqref{restr} can be just thrown away since their contributions vanish. %One can show that the same restriction occurs automatically in the space direction. 
We denote the ensemble of restricted Bratteli diagrams by $B(k,0,j,r)$, certainly $j$ should
also satisfy the restriction. %Let us denote the number of elements in $B(k,0,j,r)$ by $b(k,j,r)$.

We have a tautological formula for the density matrix:
\begin{align}
\rho=\sum_{J\in  B(2n,0,r)}\frac{Z_{-1}\{O(J)\}} {\bigl(O(J),O(J)\bigr)}\ O(J)\,.\label{density}
\end{align}
For any invariant operator $O_\mathbf{S}$
we have
$$Z_{-1}\{O_\mathbf{S}\}=\Tr_\mathbf{S}\[\rho \ O_\mathbf{S}\ q^{-2S_\mathbf{S}}\]\,.$$
Actually, this formula can be used for any, not necessarily invariant, operator: $\rho$ will project on invariant sector.
From that point of view we have a lattice analogue of Dotsenko-Fateev construction, in the scaling limit
this procedure gives the expectation values for the CFT with the central charge
\begin{align}
c=1-\frac{6\nu^2}{1-\nu}\,,\label{cc}
\end{align}
and to all the operators of the original $c=1$ CFT (scaling limit of XXZ) screening procedure is applied. 
%%%%%%%%%%%%%%%%%%%%%%%%%%%%%%%%%%%%%%%%%%%%%%%%%%%%%%%%%%%%%%%%%%%%%%%%%%%%%%%%%%%%%%%%%%%%%%%%%%%%%%%%%%%%%%%%%%%%%%%%%%%%

We are interested in the spectral properties of $\rho$. For generic $q$ the space $\(\mathbb{C}^2\)^{\otimes n}$ is 
decomposed by the irreducible representations with multiplicities:
$$
\(\mathbb{C}^2\)^{\otimes n}=\bigoplus\limits_{j=j_\mathrm{min}}^{j_\mathrm{max}}M_j\otimes V_j\,,
$$
where
$j_\mathrm{min}=0$ for $n\equiv 0\ (\mathrm{mod} 2)$, $j_\mathrm{min}=1/2$ for $n\equiv 1\ (\mathrm{mod} 2)$, 
$j_\mathrm{max}=\frac{n} 2$, and $M_j$ is the space of multiplicities whose basis is counted by $ B(n,0,j,\infty)$. 

For $q^r=1$ we use the reduced space
$$
\[\(\mathbb{C}^2\)^{\otimes n}\]_{\mathrm{rd}}=\bigoplus\limits_{j=j_\mathrm{min}}^{j_\mathrm{max}}M_{j,r}\otimes V_j\,,
$$
with $j_\mathrm{max}=\min\(\frac{n} 2,\frac{r-2} 2\)$, and the basis of $M_{j,r}$ is counted by
$ B(n,0,j,r)$. 
Being invariant the density matrix acts as unit operator on every $V_j$, and reduces effectively to the block-diagonal
matrix acting on $\oplus M_j$. Denoting blocks by $\rho_j$ we obtain

\begin{align}
\Tr\[\rho^k q^{-2S_\mathbf{S}}\]=\sum\limits_{j=j_\mathrm{min}}^{j_\mathrm{max}}\mathrm{dim}_q(j)\Tr\[\rho_j^k \]\,,\label{tracek}
\end{align}
interesting property of this formula comparing to the usual $\slt$ is that the dimensions are replaced
by quantum dimensions: traces of $q^{-2S_\mathbf{S}}$ over  irreducible representations. 

Let us consider more carefully the restricted case. Recall that the coefficients $c_{J,\al}(n)  $ have denominators
\eqref{den} which may vanish for $q^r=1$. Our first idea was that the singular ones will go when we 
restrict $J,J'$ in the main formula \eqref{main}. But the life is not so simple, actually they do not. So, working with
$q_0$ such that $q_0^r=1$ we have to take $q=q_0+\epsilon$, compute and send $\epsilon$ to $0$. Doing that
we shall first encounter singularities. Their cancelation is a property of the $\omega_{i,j}$. This is a consequence
of our general construction, but direct understanding of such property would be desirable. Then in the finite
part of the answer derivatives $\partial_\nu^k\omega_{i,j}$ appear. We shall return to this point in the next section.

Let us dwell on applications of our construction. Take a long Space of length $2N$ and consider the functional
\begin{align}
\mathbf{Z}(O)=\lim_{N\to\infty}
\frac {\Tr _\mathbf{M}\Tr_\mathbf{S}\[T_{\mathbf{S},\mathbf{M}}Oq^{-2S_\mathbf{S}}\]}
{\Tr _\mathbf{M}\Tr_\mathbf{S}\[T_{\mathbf{S},\mathbf{M}}q^{-2S_\mathbf{S}}\]}\,,\label{bfZ}
\end{align}
where $\mathbf{S}$ the tensor product of $2N$ spaces $\mathbb{C}^2$ counted from $-N+1$ to $N$. 
For $O$ being located on the interval $[1,n]$ this functional reduces to the functional $Z_{-1}$ with
$|\Psi\rangle$ being the eigenvalue of the Matsubara transfer-matrix 
$$T_\mathbf{M}=\Tr_j(T_{j,\mathbf{M}}(1)q^{-\sigma^3_j})\,,$$
with maximal eigenvalue (which we assume to be unique) since
\begin{align}\lim_{K\to\infty}\(\frac{T_\mathbf{M}}{T}\)^K=|\Psi\rangle\langle\Psi|\,.\label{Psi}\end{align}

If $|\Psi\rangle$ is quantum group invariant we are in the framework the considerations above. 
Let us show that the only possibility for $|\Psi\rangle$  to be not invariant is to be orthogonal to
the entire invariant subspace. Indeed, 
using \eqref{inv}, \eqref{Psi} one derives
$$\pi_{\mathbf{M}}(e_a)|\Psi\rangle\langle\Psi|\Phi\rangle=0\,,$$
for all $e_a$ and all invariant vectors $|\Phi\rangle$.
In some physically important cases this orthogonality can be to excluded for continuity reasons near the isotropic point $q=1$.

We can avoid all these arguments putting from the very beginning the projector on invariant subspaces $\mathcal{P}$
under the traces in \eqref{bfZ}. This gives rise to another functional
\begin{align}
\widehat{\mathbf{Z}}(O)=\lim_{N\to\infty}
\frac {\Tr _\mathbf{M}\Tr_\mathbf{S}\[T_{\mathbf{S},\mathbf{M}}Oq^{-2S_\mathbf{S}}\mathcal{P}_\mathbf{M}\]}
{\Tr _\mathbf{M}\Tr_\mathbf{S}\[T_{\mathbf{S},\mathbf{M}}q^{-2S_\mathbf{S}}\mathcal{P}_\mathbf{M}\]}\,,\label{bfZ}
\end{align}
Now the transfer-matrices would look for an invariant maximal eigenvector. 
Personally the author  does not like this violent procedure. The reasonings above show that in a normal situation $Z=\widehat{Z}$.
However, $\widehat{Z}$ is useful for mathematical rigour, in particular, as we shall see  in the next subsection.

Finally, by well-know procedure of taking limit $L\to\infty$ with staggered  inhomogeneities
\cite{klumper} one obtains from $\mathbf{Z}$ the temperature average:
$$\langle O \rangle_T=\frac{\Tr \[e^{-\frac{\mathcal{H}} T}q^{-S}O\]}{\Tr \[e^{-\frac{\mathcal{H} }T}q^{-S}\]}\,.$$

\subsection{Relation to SOS/RSOS}\label{SOS}

For invariant $O$ consider the functional $\widehat{\mathbf{Z}}(O)$ before the limit $N\to\infty$ is taken:
\begin{align}
Z_{N,L}(O)=
\frac {\Tr _\mathbf{M}\Tr_\mathbf{S}\[T_{\mathbf{S},\mathbf{M}}Oq^{-2S_\mathbf{S}}\mathcal{P}_\mathbf{M}\]}
{\Tr _\mathbf{M}\Tr_\mathbf{S}\[T_{\mathbf{S},\mathbf{M}}q^{-2S_\mathbf{S}}\mathcal{P}_\mathbf{M}\]}\,.\nn
\end{align}
We have seen that for $q^r=1$ the restriction $j\le(r-2)/2$ for the intermediate spins is automatic.
Let us try to rewrite $Z_{N,L}(O)$ in terms of SOS for generic $q$ or RSOS for $q^r=1$
configurations. 

The SOS configurations are counted by spins $j=0,1/2,1,\cdots$ put
on the faces of a square lattice, for RSOS case they are restricted. The spins are subject to the restriction $j'-j=\pm 1/2$ across vertical and horizontal edges. Folklore says that the SOS model is equivalent to the six-vertex model. But what does it mean exactly?
On a torus, which is considered here,  the number of six vertex configurations is finite while the number of SOS configurations
is infinite. Let us consider the equivalence carefully rewriting the functional $\widehat{Z}$ in the SOS/RSOS language.

We start by the partition function in the denominator of $Z_{N,L}$. The vertices $(i,j)$ are counted by space coordinate $i$ and Matsubara coordinate $j$.
We count a faces by its left-lower vertex. 
 We have seen that $T_\mathbf{M}$ acts from the invariant space to itself, so,
 The projector $\mathcal{P}$ can be freely moved along the space direction.
 Let us put it between the $0$-th and $1$-st lines. 
 The natural orthogonal basis in the invariant 
 subspace follows from Section \ref{QG}, it consists of
 $E_{L,J,0,0}$, $j_0=j_k=0$, we construct our projector out of them.
 One can rewrite $T_\mathbf{M}$ in the basis $E_{L,J,0,0}$, however, combining  $(T_\mathbf{M})^N$ we do not obtain
 a local SOS interaction: all the spins on the faces $(\star,0)$ equal $0$, so, they do not satisfy the SOS condition
 along the space direction.

We shall avoid this non-locality problem using that the vectors
 $$E_{L,J}=\sum_{l=0}^{2j}q^{2(-j+l)}E_{L,J,-j+l,-j+l}\,,\quad j_1=j_{2n+1}=j\,,\quad  j_j=j_k=j\,,$$
which  also happen to be invariant
(for $E_{L,J,0,0}$ redundant indices are omitted).
For $j_0>0$ these vectors are {\it not linearly independent},
this circumstance will cause some non-locality on the border. 

Let us renormalise these vectors
$$|J\rangle=\frac 1 {\prod_{p=1}^{L-1} {\sqrt{2j_p+1}}}\ E_{L,J}\,,$$
in order that for $j_0=0$ we have an orthonormal basis. 
Since for $q^r=1$ the intermediate spins are restricted we shall not meet zeros in the denominators.

We observe 
that the transfer-matrix $T_\mathbf{M}$ acts locally on $E_{J}$:
\begin{align}
T_\mathbf{M}|J\rangle=\sum_{J':\ j_p'-j_p=\pm 1/2}X_{J,J'}|J'\rangle\,,\label{move}
\end{align}
where
$$X_{J,J'}=\prod\limits_{p=0}^{L-1}F(t_p^{-1},j'_p,j_p,j'_{p+1},j_{p+1})\,,$$
with the coefficients $F$ obtained using the $6j$-coefficients. Explicitly
\begin{align}
&F(x,j,j+1/2,j+1/2,j+1)=\frac{\sqrt{[2j+1]}}{\sqrt{[2j+2]}}(qx-q^{-1}x^{-1})\,\nn\\
&F(x,j,j-1/2,j-1/2,j+1)=\frac{\sqrt{[2j+1]}}{\sqrt{[2j]}}(qx-q^{-1}x^{-1})\,,\nn\\
&F(x,j,j+1/2,j+1/2,j)=\frac1{\sqrt{[2j+1]}\sqrt{[2j+2]}}(x^{-1}q^{2j+1}-xq^{-(2j+1)})\,,\nn\\
&F(x,j,j-1/2,j-1/2,j)=\frac1{\sqrt{[2j+1]}\sqrt{[2j]}}(q^{2j+1}x-q^{-(2j+1)}x^{-1})\,,\nn\\
&F(x,j,j+1/2,j-1/2,j)=\frac{\sqrt{[2j+2]}}{\sqrt{[2j+1]}}(x-x^{-1})\,,\nn\\
&F(x,j,j-1/2,j+1/2,j)=\frac{\sqrt{[2j]}}{\sqrt{[2j+1]}}(x-x^{-1})\,,\nn
\end{align}
the parameters $t_p$ are the inhomogeneities of the Matsubara chain.

Let us present the identity \eqref{move} graphically. 
We use Kirillov-Reshetikhin notations \cite{KR}, namely,
for $R$-matrix we use cross of solid lines,
for the face $R$-matrix $F$ we use cross of dashed lines. The vector $|J\rangle$ is represented by
closed dashed line with solid intervals pointed to the left, the initial point is marked by a short
fat interval, finally, bullet in the left hand side stands for the insertion of $q^{-\sigma^3}$.
\vskip .2cm
\centerline{\includegraphics[height=6cm]{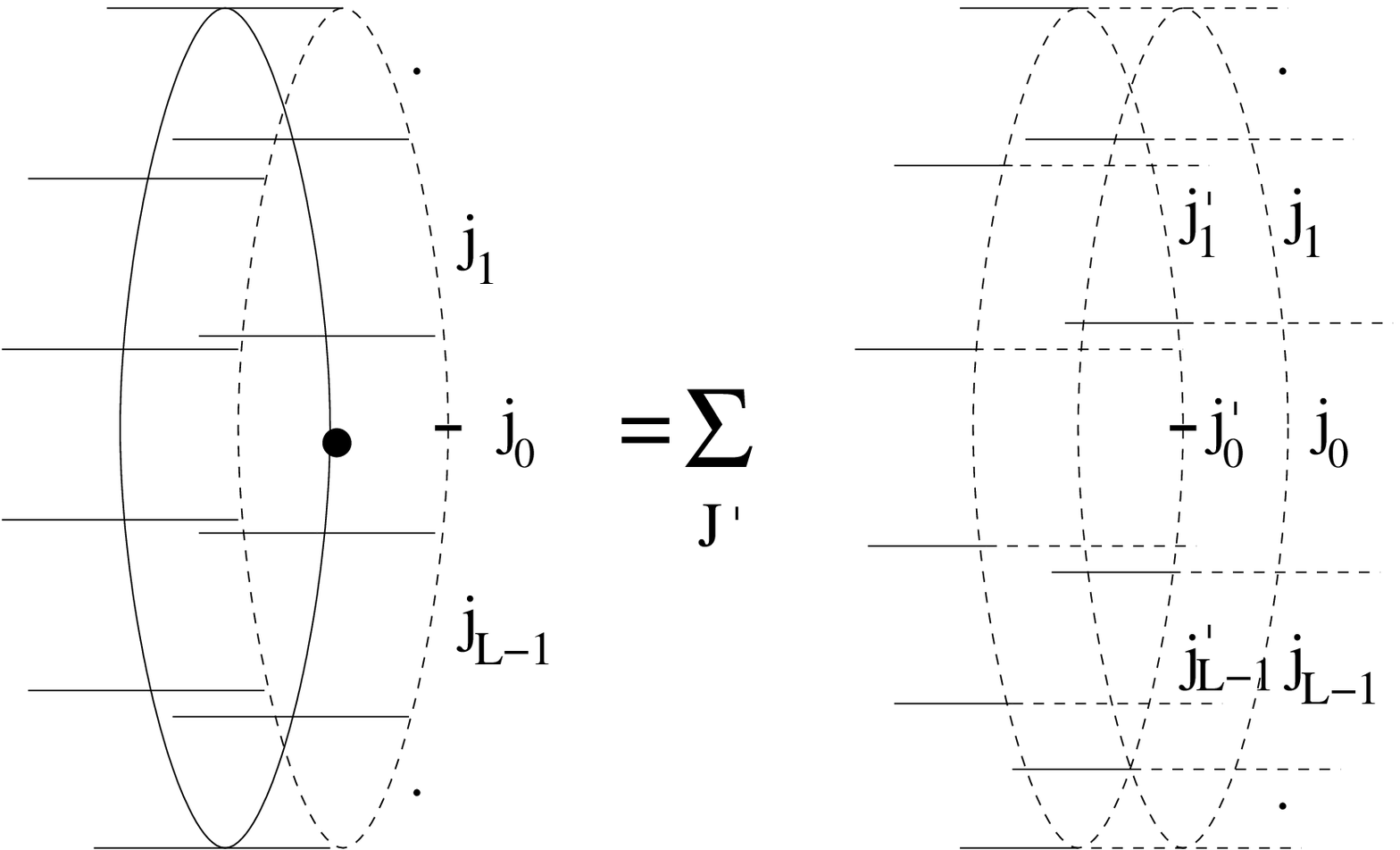}}
{\it fig.2} First step of vertex-face correspondence. 
\vskip .2cm
Already at this point we see that the equivalence of the invariant sector of XXZ and
SOS/RSOS is far from being clear: in SOS/RSOS formulation different
face configurations are supposed to be independent while from XXZ prospective
they correspond to linearly dependent vectors.

Using this and similar identities we obtain for $T_{\mathbf{S},\mathbf{M}}Oq^{-2S_\mathbf{S}}\mathcal{P}_\mathbf{M}$ which enters the numerator of $Z_{N,L}$ the equality
presented graphically on {\it fig.3}. For $J=\emptyset$ we get
$T_{\mathbf{S},\mathbf{M}}q^{-2S_\mathbf{S}}\mathcal{P}_\mathbf{M}$ present in the denominator. 
\vskip .2cm
\centerline{\includegraphics[height=6cm]{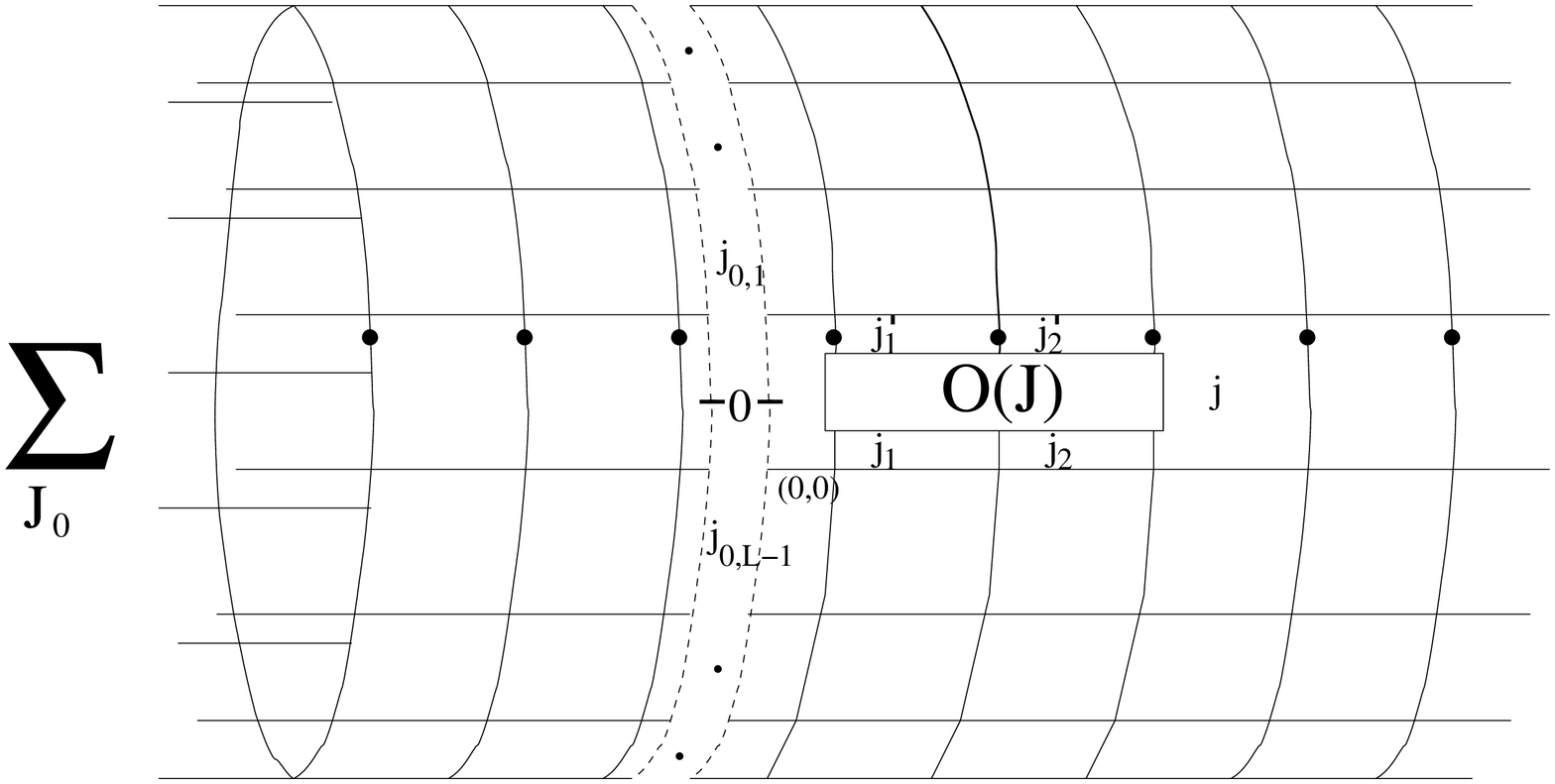}}
\vskip.8cm
\centerline{\includegraphics[height=6cm]{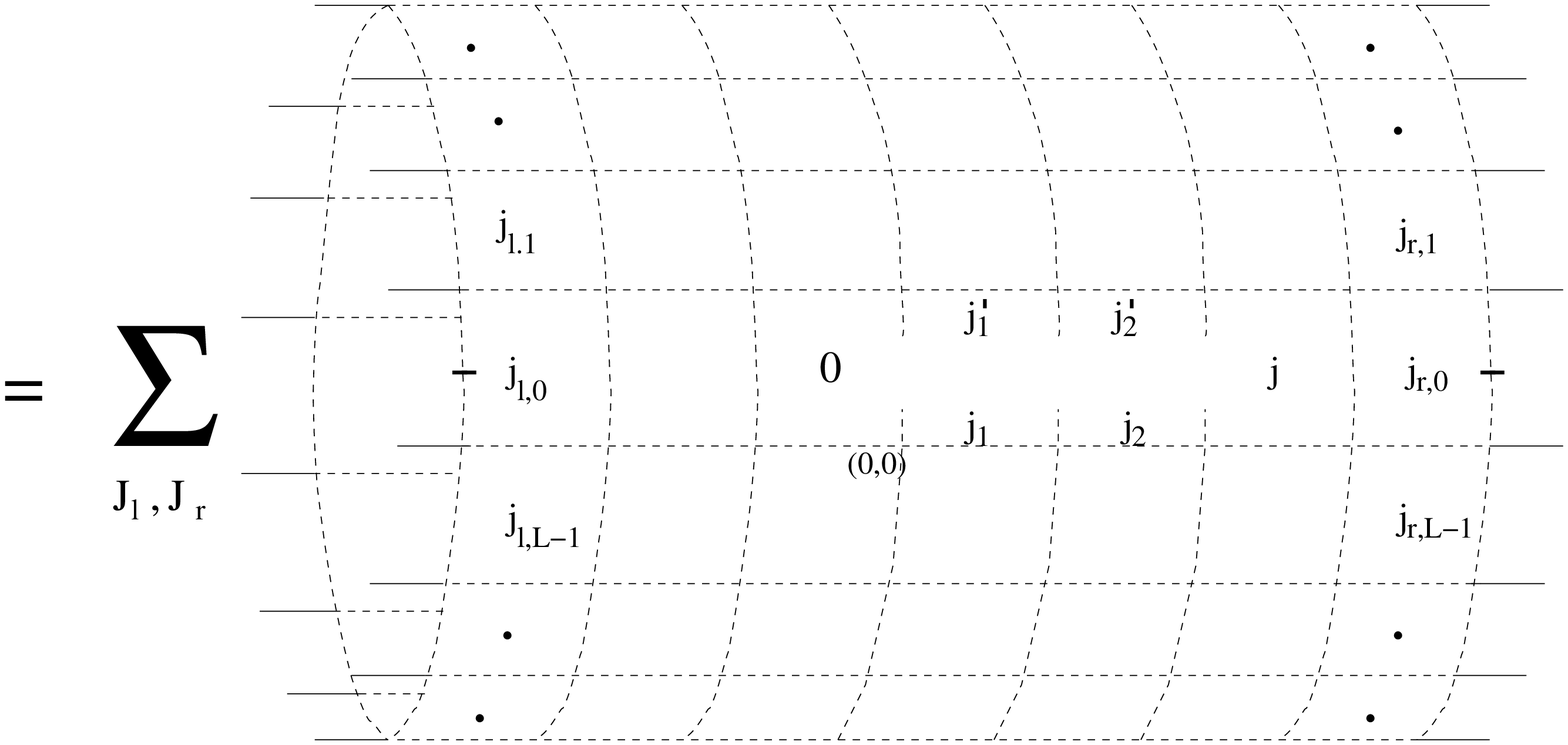}}
{\it fig.3} Global vertex-face correspondence.
\vskip .2cm

Now it is clear that the functional $Z_{N,L}(O)$ can be rewritten in SOS/RSOS formulation,
but if we consider different face configurations as independent and take trace
over them then we have to insert under this trace an operator $M$ whose
matrix elements are
$$M_{J,J'}=\langle J|J'\rangle\,.$$
In our opinion even in the limit $N\to \infty$ this boundary contribution cannot be dropped except, probably,
for such rough characteristics of the system as the free energy. 

\section{Numerical results. Zero temperature}

In the zero temperature limit the twist $q^{-2S}$ becomes irrelevant, and the
Bethe roots for the maximal eigenvector are real, dense, distributed with Lieb density. 
For the isotropic case the maximal eigenvector $|\Psi\rangle$ is $\slt$-invariant. The us take
some simple invariant vector $|\Phi\rangle $ to which it is not orthogonal (for example one from
the second basis of the Subsection \ref{IO} basis). The scalar product 
\begin{align}f(\nu)=\langle\Psi|\Phi\rangle\,.\label{scpr}\end{align}
is an analytical function of $\nu$. There is no reason to believe that $f(\nu)$ has branch points
on the interval $([0,1)]$. So, as analytical function which is not identically $0$ the function
$f(\nu)$ has a discreet set of zeros on the interval.

In order to write down the function $\omega(x,y)$ it is convenient to make a change of variables :
 $$\la=\frac 1 {2\pi\nu}\log x\,.$$ 
 We shall slightly abuse notations using the same letters for functions of different variables:
 $$\psi(\la)=\pi\nu\tanh(\pi \nu\la)\,,\quad K(\la)=\frac 1 {2\pi i}\(\psi(\la+i)-\psi(\la-i)\)\,,$$
 but we shall change the notation for $\omega$:
$$ \varpi(\la,\mu)=\omega \bigl(e^{2\pi\nu\la},e^{2\pi\nu\mu}\bigr) \,.$$

The function $\varpi(\la,\mu)$ is simple for $T=0$:
\begin{align}
&\varpi(\la,\mu)=\varpi_0(\la-\mu)\,,\label{omega0}\\
&\varpi(\la)=\int_{-\infty}^{\infty}\cos (2\la k )\frac{4\sinh\(\frac{(1-\nu)}\nu k\)}
{ \sinh\(\frac 1\nu k \)\cosh( k)}dk
-\frac{\pi } 2 K(\la)\,.\nn
\end{align}
It is not difficult to recalculate the  Taylor series for $\varpi(\la,\mu)$ in $\la,\mu$ as the Taylor series 
for $\omega(x,y)$ in $x-1,y-1$. 

We consider the values of $\nu$ which correspond in the scaling limit to minimal models $M_{r,s}$ with
$s\le 11$.
The relation to $\nu$ is
$$\frac r s=1-\nu\,.$$

The procedure is simple: we take $\omega(x,y)$ given by \eqref{omega0} plug it into \eqref{main}, compute the density
matrix and diagonalise it block-wise. The only problem is that for $s\le 7$ there are singularities in $c_{J,\al}$. So we
check that they disappear in the final formula, and the final result contains not only the function $\omega(x,y)$, but also
its derivatives with respect to $\nu$ which are not hard to compute, just a little boring. 

What do we do having the spectra of the density matrix? First thing to do is to check that the
invariant trace of $\rho$ equals $1$.
For all $\nu$'s under consideration except $\nu=\frac {s-1} {s}$ this is true.
Let us give a simple  example 
$$n=2,\quad \nu=\frac2{11}\,,$$
Anticipating the results of the next section we consider different temperatures.
The density matrix has only two $1\times 1 $ blocks with $j=0,1$,
let $\rho_0$ and $\rho_1$ are the corresponding  eigenvalues.
In the table below we set the accuracy $10^{-24}$, but actually we compute with 60 digits.
$$
\begin{tabular}{|c|c|c|}\hline$T$&$\rho_0$&$\rho_1$\\ \hline
0 &0.749590171881005195039534  &0.136773935626112993010667 \\ \hline
1/16 &0.749312246998088632583255   &0.136925738294284871532951\\ \hline
1/8 &0.748472992599050986722480 &0.137384139340322728234512\\ \hline
3/16 & 0.747055192673494195413593&0.138158542155122243436699\\ \hline
1/4 &0.745025105413502060299643 & 0.139267376525958562486133\\ \hline
\end{tabular}$$
With these data one checks that always
$$\rho_0+\frac{\sin\(\frac {6\pi}{11}\)}{\sin\(\frac {2\pi}{11}\)}\rho_1=1\,.$$

For $\nu=\frac {s-1} {s}$ which correspond to degenerate minimal models $M_{1,s}$ we get
unreasonable result: the invariant trace of $\rho$ is not equal to $1$.
Presumably these are the points where the scalar product
\eqref{scpr} vanishes for all $|\Phi\rangle$?

 All together we consider
$$ \nu\in\left\{\frac{1}{11},\frac{ 1}{10},\frac{ 1}{8},\frac{ 1}{7},\frac{ 1}{6},\frac{ 2}{11},\frac{ 1}{5},\frac{ 1}{4},\frac{ 3}{11},\frac{ 2}{7},\frac{ 3}{10},\frac{ 4}{11},\frac{ 3}{8},\frac{ 2}{5},\frac{ 3}{7},\frac{ 5}{11},\frac{ 6}{11},\frac{ 4}{7},\frac{ 3}{5},\frac{ 5}{8},\frac{ 7}{11},\frac{ 11}{16},\frac{ 7}{10},\frac{ 5}{
   7},\frac{ 8}{11},\frac{ 7}{9}\right\}\,.$$
There is one additional point $\nu=11/16$ which we included in order to see more
clear what happens near the point $\nu=3/4$.

Now we can compute $\Tr\[\rho ^k q^{-2S}\]$. For integer $k$ we observe a combination of
monotonous and staggering behaviours, making it hard to compare with the scaling limit with data at hand.
But for $k\ll 1$ the behaviour gets smoother and becomes really nice for the Von Neumann entropy.
The latter is defined as 
\begin{align}s(\nu,n)=-\sum\limits_{j=j_\mathrm{min}}^{j_\mathrm{max}}\mathrm{dim}_q(j)\Tr\[\rho_j\log(\rho_j) \]\,,
\label{vN}\end{align}
where $\rho$ is computed on $n$ sites. 

In the scaling $n\to \infty$ limit $s(\nu,n)$ should behave as $c/3\log n +B$, with $B$ being some non-universal constant.
With our data we find typically pictures like that
\vskip.4cm
%\begin{figure}
\centerline{\includegraphics[height=8cm]{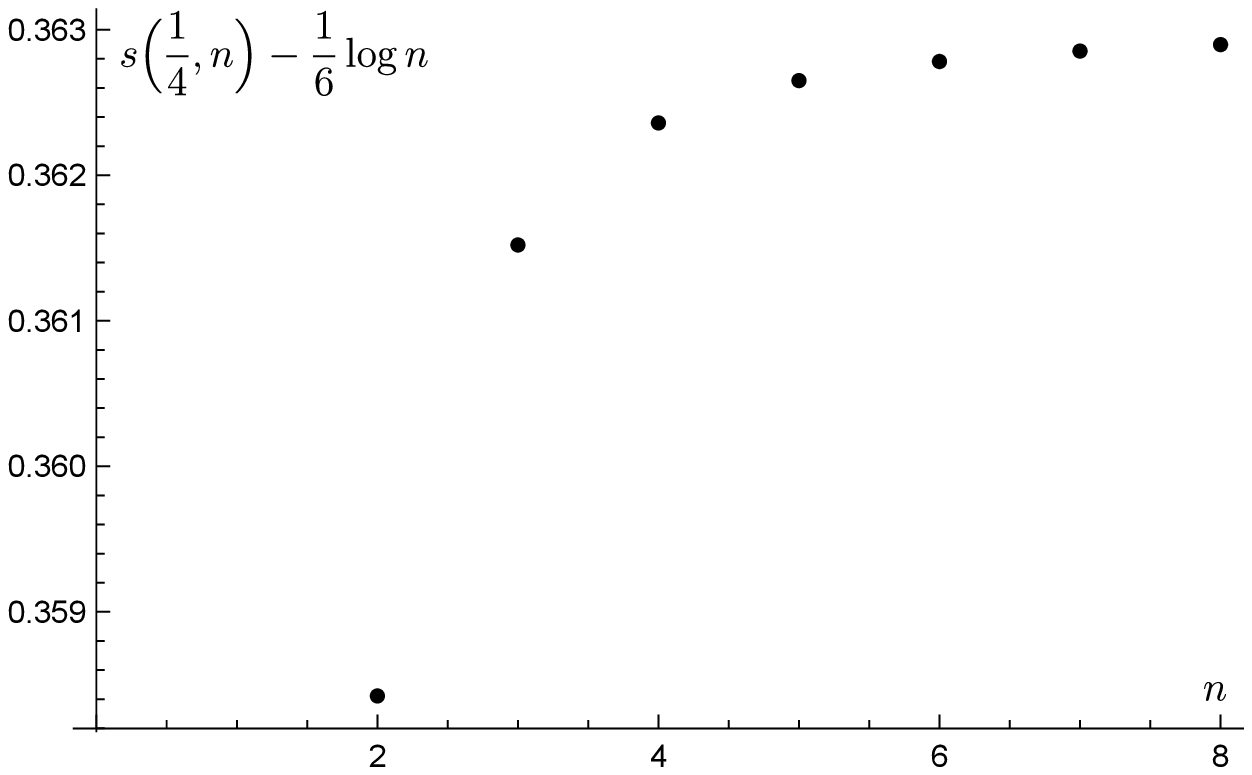}}
{\it fig.4 } Entropies for $\nu=1/4,\ T=0$.%\end{figure}
\vskip.2cm
Analysing these data we come to the conclusion that the first correction to the CFT behaviour
is $O(n^{-2})$.
Then we fit the numerical data for the largest $n$'s at hand, $n=6,7,8$, with the ansatz
$$s(\nu,n)=\frac{c(\nu)} 3\(A(\nu)\log n-G(\nu)n^{-2}\)+B(\nu)\,.$$
Our results agree with CFT if $A(\nu)$ is close to $1$ and $G(\nu)$ is sufficiently small. 
The correction $O(n^{-2})$ is experimental. It should come from two places: the contribution of
the descendants in the definition of operators in CFT perturbation, and from the perturbation of the
CFT Hamiltonian \cite{lukyanov}. 

Here are our results for $A(\nu)$. We see that they are generally good, and some times amazingly good having
in mind relatively small sizes of our sub-lattices.
\vskip .2cm

\centerline{\begin{tabular}{|c|c|c|c|c|c|c|c|c|c|c|c|c|}
  \hline
$\,  \ \nu\ \  \, $& 1/11& 1/10& 1/8& 1/7& 1/6& 2/11& 1/5& 1/4
\\ \hline$\, \ \ A(\nu)\ \ \, $&1.00103& 1.00081& 1.00032& 1.00007& 0.99984& 0.99974& 0.99966& 
0.99959\\\hline
   \end{tabular}}
\vskip .5cm
\centerline{\begin{tabular}{|c|c|c|c|c|c|c|c|c|c|c|c|c|}
  \hline
3/11& 2/7& 3/10& 4/11& 3/8& 2/5& 3/7& 5/11& 6/11
\\ \hline0.99959& 0.99959& 0.99960& 0.99962& 0.99962& 0.99963& 0.99965& 
0.99966& 0.99977\\\hline
   \end{tabular}}

\vskip .5cm
\centerline{\begin{tabular}{|c|c|c|c|c|c|c|c|c|c|c|c|c|}
  \hline
4/7& 3/5& 5/8& 7/11& 11/16& 7/10& 5/7& 8/11& 7/9
\\ \hline0.99982& 0.99990& 0.99999& 1.00004& 1.00030& 1.00037& 1.00044& 
1.00047& 1.00962\\\hline
   \end{tabular}}
\vskip .2cm
The worst agreement is close to the isotropic case $\nu=0$. 

The constant $B(\nu)$ for $1/2<\nu<2/3$ acquires an imaginary part which equals  $\pi$
with a great precision. So, it makes sense to consider $e^{B(\nu)}$. Here is the graphics 
with the last point, $\nu=7/9$ omitted. The latter belongs to the interval $[3/4,4,5]$, we do not have enough
enough points in this interval to draw conclusions.

\vskip.2cm
%\begin{figure}
\centerline{\includegraphics[height=8cm]{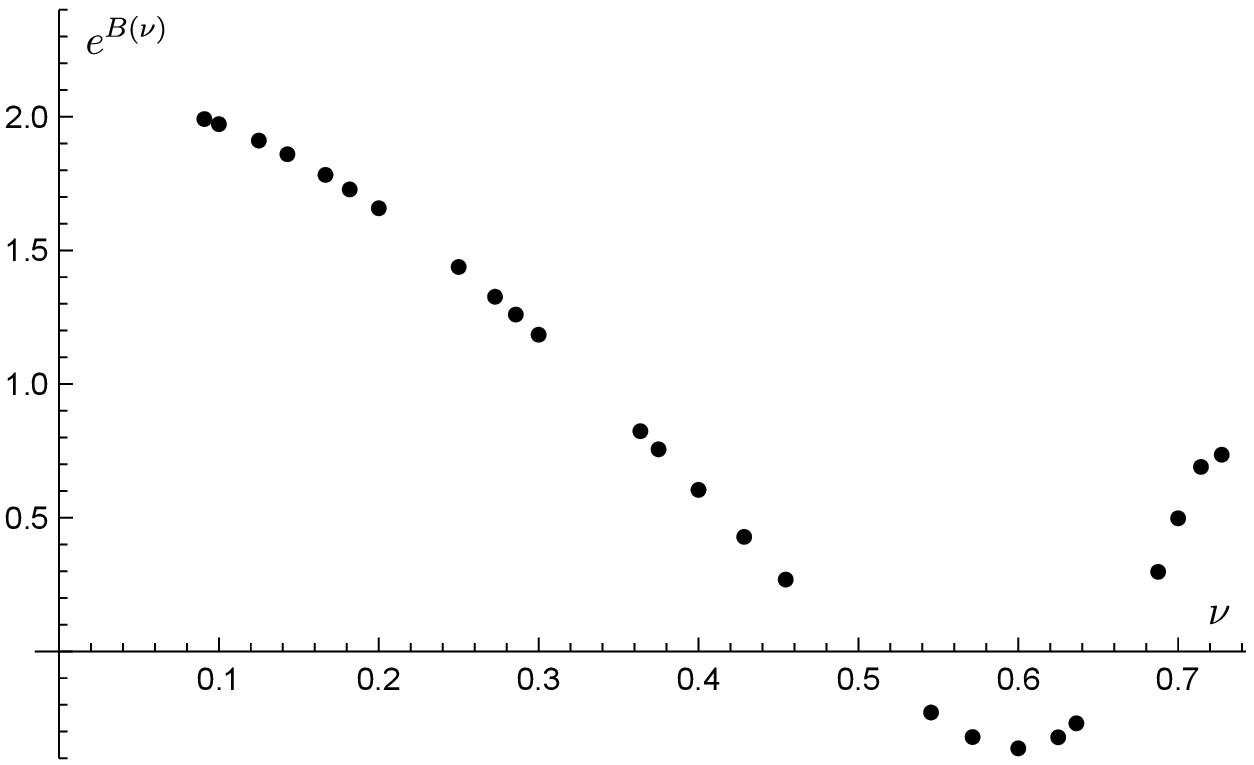}}
{\it fig.5} Behaviour of $e^B$ as function of $\nu$.%\end{figure}
\vskip.2cm

Finally, for the constant $B(\nu)$ we obtain the following ( $\nu=7/9$ is again omitted).
\vskip.2cm
%\begin{figure}
\centerline{\includegraphics[height=8cm]{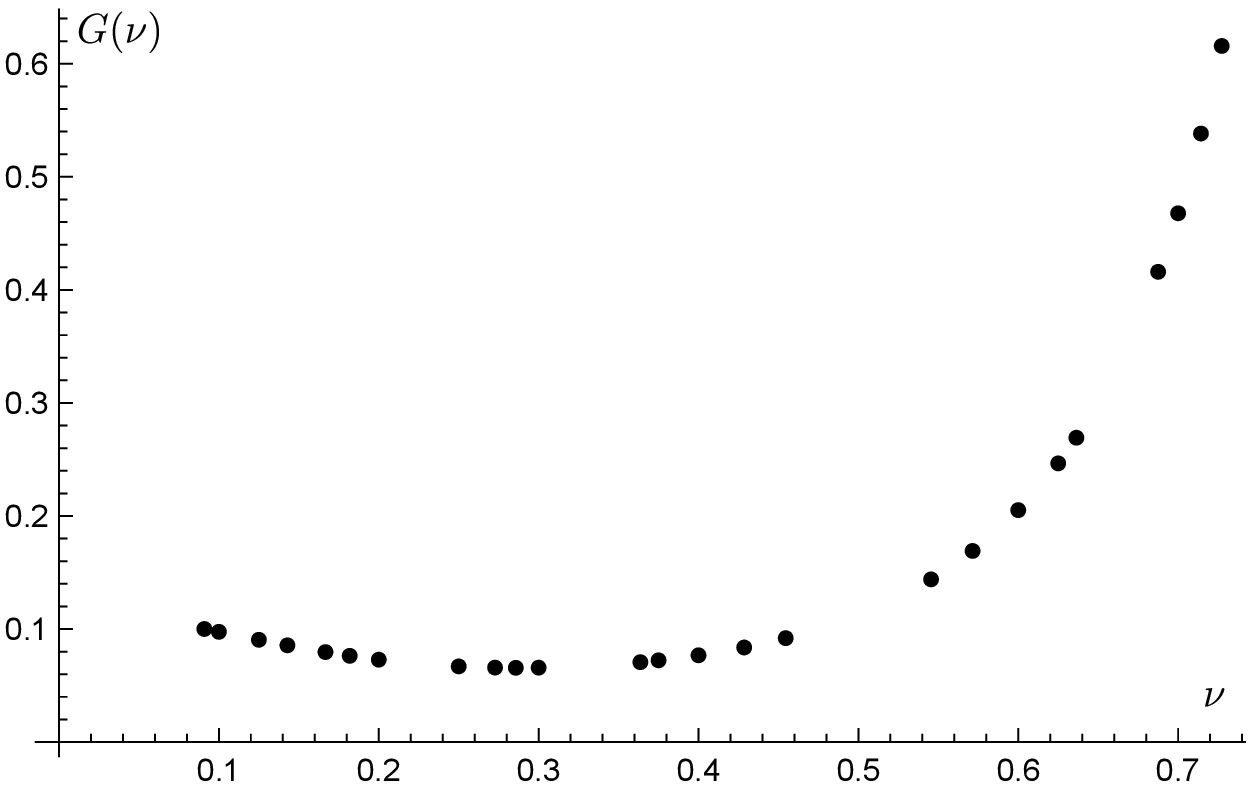}}
{\it fig.6} Behaviour of $G$ as function of $\nu$%\end{figure}
\vskip.2cm

Trying to fit data for  $\nu=9/11$
we get unreasonable results. 
Let us see what happens. First, $s(9/11,n)$ has an imaginary part. However,
the imaginary part is getting relatively small with $n$ growing, here are the arguments
of $-s(9/11,n)$ for for $n=6,7,8$ respectively: $-0.0131820, -0.0006312,-0.000017
$. For the real part we have
\vskip.2cm
%\begin{figure}
\centerline{\includegraphics[height=8cm]{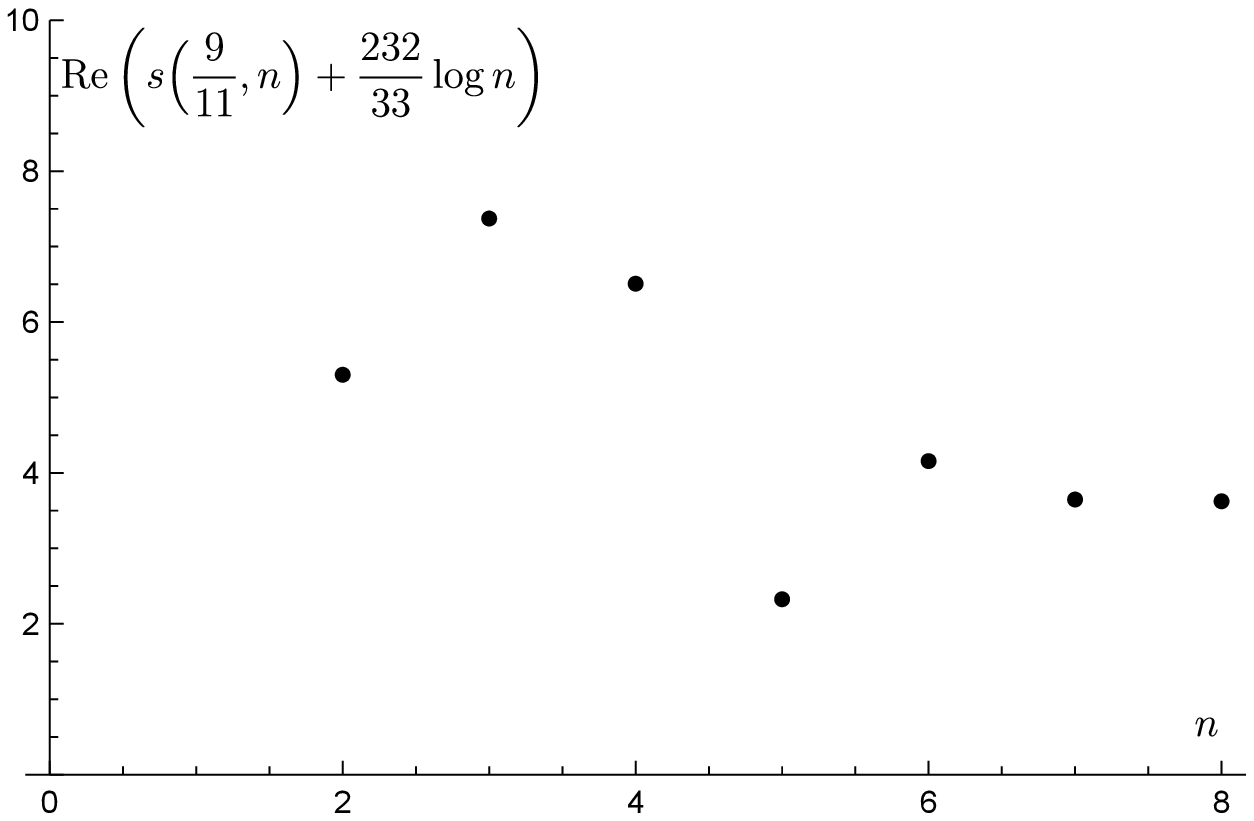}}
{\it fig.7 }  Entropies for $\nu=9/11,\ T=0$.%\end{figure}
\vskip.2cm

This is quite different form the typical behaviour on {\it fig}. For small $n$ the behaviour is complicated, only the last three points seem to start approaching the CFT behaviour. 
However, more careful look convinces that the point $n=6$ is still rather far from
the asymptotical  trend. It would be desirable to have results for $n=9$, but this is hard
for the moment. 
\section{Temperature}

Computation of the function $\varpi(\la,\mu)$ for the case of finite temperature
is not very different from the isotropic case  \cite{ms} that is why we shall be brief. 
We shall consider $\nu$ in the range from $0$ to $1/2$. Above the latter, free fermion, point
precision of our numerics drops drastically requiring more work.

We assume that in the ground state all Bethe roots are real and situated inside the interval $\la\in [-R,R]$.
The value of $R$ must be consistent with the requirement
\begin{align}|\log(\mathfrak{a}(\pm R))|<\pi,\,,\label{aconditions}
\end{align}
where the counting function $\mathfrak{a}(\la)$ will be defined soon. 
We have 
$$|\mathfrak{a}(\la )|=1,\quad \la\in \mathbb{R}\,,$$
the Swartz principle asserts 
$$ \mathfrak{a}(\bar \la )=\frac 1 {\overline{\mathfrak{a}( \la )}}\,.
$$

We have to define several objects. The first one is the resolvent $R(\la,\mu)$ which solves the equation
\begin{align}R(\la,\mu)=K(\la-\mu)+\int\limits_{-R}^{R}K(\la-\eta)R(\eta,\mu)d\eta\,.\label{eqR}\end{align}
The resolvent is continued analytically by virtue of the equation \eqref{eqR} itself. 
We consider the ellipse
\begin{align}\la(\phi)=-R\cos(\phi)-it\sin\phi, \quad 0\le \phi<2\pi\,.\label{ell}\end{align}
The parameter $t$ must be between $0$ and $1$. In numerical solution we usually take $t=2/5$. 
We denote by $C_\pm$ the parts of the ellipse belonging to $\mathbb{C}^{\pm}$, both oriented from left to right.

We shall need the function
\begin{align}
F(\la,\mu)=f(\la-\mu)+\int\limits_{C_+}R(\la,\eta)f(\eta-\mu)d\eta,\qquad f(\la)=\psi(\la-i)-\psi(\la)\,.\nn
\end{align}
In order to avoid singularities situated close to the integration contour this equation is rewritten as
\begin{align}
&F(\la,\mu)=F_0(\la-\mu)+\Delta F(\la,\mu)\,,\nn\\
&\Delta F(\la,\mu)=d(\la,\mu)+\int\limits_{-R}^RR(\la,\eta)d(\eta,\mu)d\eta\,,\nn\\ 
&
 d(\la,\mu)=-\Bigl(\int\limits_{-\infty}^{-R}+\int\limits_{R}^{\infty}\Bigr)K(\la-\eta)F_0(\eta-\mu)d\eta\,.\nn
\end{align}

Then we have the NLIE:
\begin{align}
\log\mathfrak{a}(\la)=\frac 1 T {F(\la,0)}  -\int\limits_{C_+}R(\la,\mu)\log\( 1+\mathfrak{a}(\mu)\)d\mu+
\int\limits_{C_-}R(\la,\mu)\log\( 1+\overline{\mathfrak{a}(\mu)}\)d\mu\,.\label{neweqa}
\end{align}
The iterations for this equation converge fast because $|\mathfrak{a}(\mu)|\ll 1$ for $\mathrm{Re}(\mu)$ close
to $0$ (at this point $\mathfrak{a}(\mu)$ develops an essential singularity.

Introduce the measures
$$dm(\la)=\frac {d\la}{1+\mathfrak{a}(\la)}\,,\quad
d\overline{m}(\la)=\frac {\mathfrak{a}(\la)d\la}{1+\mathfrak{a}(\la)}\,.$$
We need the auxiliary $G(\la,\mu)$ function which solves the equation
\begin{align}
G(\la,\mu)=F(\la,\mu)-\int\limits_{C_+}R(\la,\eta)G(\eta,\mu)d\overline{m}(\eta)-\int\limits_{C_-}R(\la,\eta)G(\eta,\mu)dm(\eta)\,.\nn
\end{align}
Now we are ready to define the function $\varpi(\la,\mu)$:
\begin{align}
&\varpi(\la,\mu)=\varpi_0(\la-\mu)+{\varpi_1(\la,\mu)}+\varpi_2(\la,\mu)\,,\label{omega}\\
&\varpi_1(\la,\mu)=-\frac 1 {2\pi}\Bigl(\int\limits_{-\infty}^{-R}+\int\limits_{R}^{\infty}\Bigr)
f(\la-\eta)F_0(\eta-\mu)d\eta+\frac 1 {2\pi}\int\limits_{C_-}f(\la-\eta)\Delta F(\eta,\mu)d\eta\,,\nn\\
&\varpi_2(\la,\mu)=\frac 1 \pi\Bigl(\int\limits_{C_+}F(\la,\eta)G(\eta,\mu)d\overline{m}(\eta)
+\int\limits_{C_-}F(\la,\eta)G(\eta,\mu)dm(z)\Bigr)\,.\nn
\end{align}

For small $T$ the functions $\varpi_1(\la,\mu)$, $\varpi_2(\la,\mu)$ are small corrections to $\varpi_0(\la-\mu)$, and we get rather fast numerical procedure, parallel to that of \cite{ms}. We compute up to $T$ from $1/160$ to
$1/4$. 
Smaller is $T$ greater interval we need. To be precise we  use $R=1,2,3$. Certainly, we could do everything with
$R=3$, but it requires more precision and more time consuming. For numerical integration we use the double exponential 
method \cite{MS} approximating \cite{ooura}
\begin{align}
&\int_{-R}^{R}f(x)dx\simeq hR\sum_{k=-N}^{N}f(Rg(hk))g'(hk)\,.\nn\\
&g(t)=-1+\frac 4 \pi \arctan(\exp(c\sinh(t)))\,.\nn
\end{align}
then for different $R$ we set
\begin{align}
&h=1/20,\ \ N=200,\quad \mathrm{for}\ \ R=1\,,\nn\\
&h=1/25,\ \ N=250,\quad \mathrm{for}\ \ R=2\,,\nn\\
&h=1/40,\ \ N=400,\quad \mathrm{for}\ \ R=3\,.\nn
\end{align}
for integrals over real axis and
\begin{align}
&h=1/30,\ \ N=300,\quad\mathrm{for}\ \ R=1\nn\\
&h=1/40,\ \ N=400,\quad \mathrm{for}\ \ R=2\,,\nn\\
&h=1/60,\ \ N=600,\quad \mathrm{for}\ \ R=3\,.\nn
\end{align}
for integration over a  half of the ellipse. 

For the finite temperature case we consider only the most advanced case $n=8$.
We want to avoid the complications related to singularities of $c_{J,\al}(n)$ because computing derivative
with respect to $\nu$ of functions defined above present additional, unnecessary, problem. So, we take $\nu$
with denominator $11$. Also we work only below the free fermion point: $\nu<1/2$, so, we take
$$n=1/11,\ 2/11,\ 3/11, \ 4/11,,\ 5/11\,,$$
which corresponds to the minimal models $M_{r,11}$, $r=1,2,3,4,5$. This is quite representative because we have
in this list the unitary model $M_{1,11}$ and non-unitary models with positive and negative central charges. 

The CFT predicts that the difference between the Von Neumann entropies at temperature $T$ and at zero
temperature behaves as
$$s(n,T)-s(n,0)=\frac {c} 3 \log\(\frac{\sinh (nT)}{nT}\)\,.$$
The following figure presents comparing of our results with the CFT prediction up to $8T=2$.
We see that agreement is good for the unitary case, and is getting better when we move further
from the isotropic point. 
%\centerline{\includegraphics{pf.eps}}

\newpage
\vskip.2cm
%\begin{figure}
\centerline{\includegraphics[height=15cm]{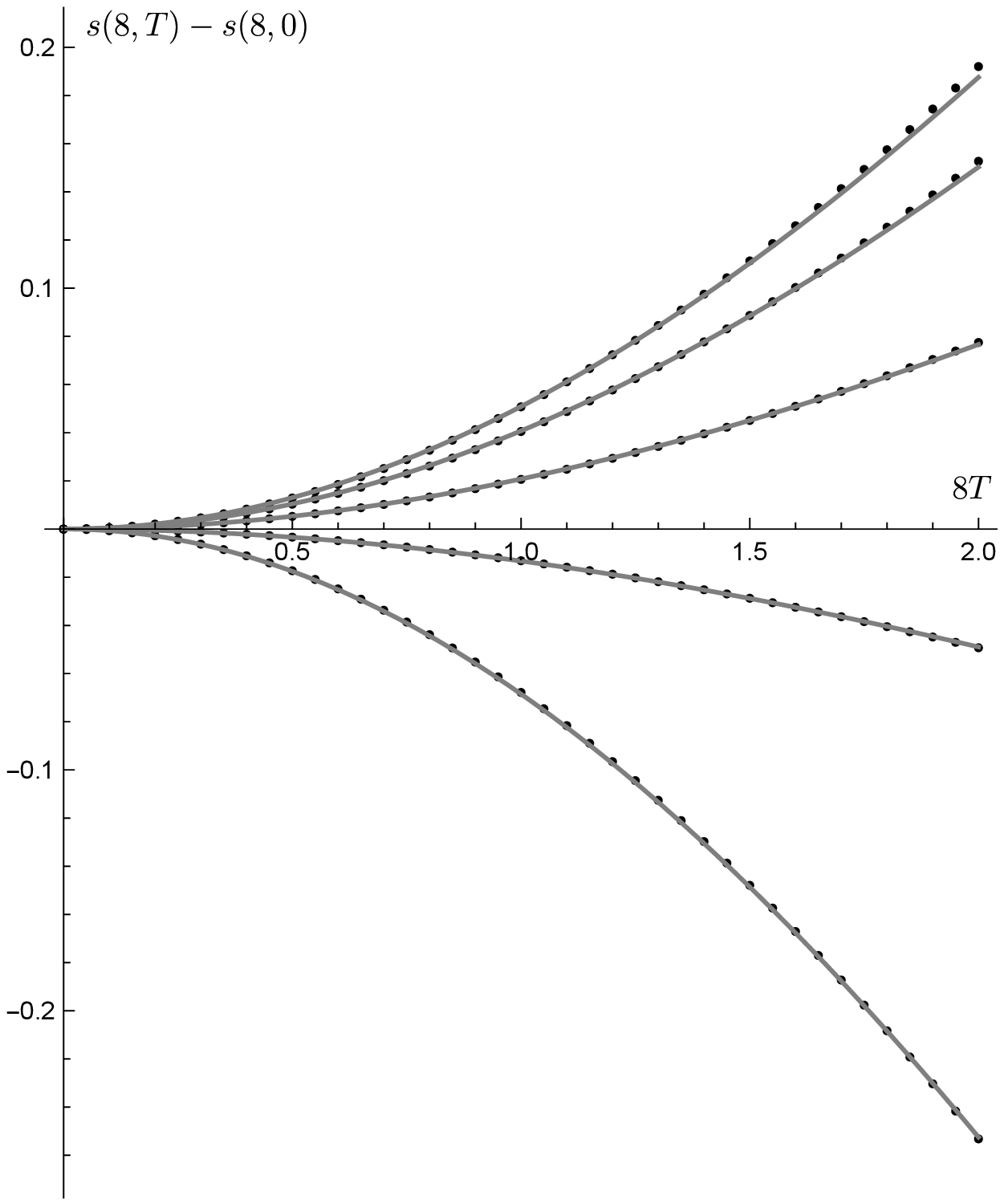}}
{\it fig.8 } Comparison with the minimal models $M_{p,11}$ for $p=1,2,3,4,5$ (from up to down).
%\end{figure}
\vskip.2cm
We think that this beautiful picture is a good point to finish this paper. 

{\bf Acknowledgements} For the case of finite temperature I used slightly modified version of
a program written together with T. Miwa, I am greateful to him for his effort. I also acknowledge 
discussions with B. Estienne, Y. Ikhlef  and N. Reshetikhin.

 \end{document}